\documentclass[%
 reprint,
 amsmath,amssymb,
aps,
pra,
]{revtex4-2}

\usepackage{graphicx}                            %
\usepackage[a4paper, total={6in, 8in}]{geometry} %
\usepackage{xcolor}                              %
\usepackage{subcaption}                          %
\usepackage{amsmath}        	                 %
\usepackage{appendix}                            %
\usepackage{bm}
\usepackage{tabularray}                           %
\usepackage{comment}
\usepackage[export]{adjustbox}   %

\usepackage{tikz}
\usetikzlibrary{intersections}

\newcommand{\cyldeform}{\delta} %
\newcommand{\re}{\mathrm{Re}\,}
\newcommand{\im}{\mathrm{Im}\,}
\renewcommand{\Re}{\mathrm{Re}\,}
\renewcommand{\Im}{\mathrm{Im}\,}
\newcommand{\imag}{i}
\newcommand{\euler}{\mathrm{e}}
\renewcommand{\vec}[1]{\bm{#1}}
\newcommand{\speedoflight}{\mathrm{c}_0}

\newcommand{\distancefunc}{\mathcal{L}}
\newcommand{\phasefunc}{\mathcal{P}}
\newcommand{\geodesic}{\vec{g}}
\newcommand{\geodesiclength}{\geodesic_\ell}
\newcommand{\geodesicphase}{\geodesic_\mathrm{p}}
\newcommand{\geodesicortho}{\geodesic_\mathrm{o}}

\newcommand{\qnphi}{m}     %
\newcommand{\qnrr}{\rho}   %
\newcommand{\qnzz}{\eta}    %
\newcommand{\hamiltonian}{\mathcal{H}}

\newcommand{\spatialwidth}{8em}

\begin{document}

\title{
Exceptional-point-controlled mode interaction in three-dimensional microcavities represented by generalized Husimi functions
}

\author{Tom Simon Rodemund}
\affiliation{
Institute of Physics, Technische Universität Chemnitz, D-09107 Chemnitz, Germany
}
\affiliation{
Light-Matter Interactions for Quantum Technologies Unit, Okinawa Institute of Science and Technology Graduate University,
Onna, Okinawa 904-0495, Japan
}

\author{Shilong Li}
\affiliation{
Interdisciplinary Center for Quantum Information, State Key Laboratory of Modern Optical Instrumentation,
College of Information Science and Electronic Engineering, Zhejiang University, Hangzhou 310027, China
}
\affiliation{
Light-Matter Interactions for Quantum Technologies Unit, Okinawa Institute of Science and Technology Graduate University,
Onna, Okinawa 904-0495, Japan
}

\author{Síle {Nic Chormaic}}
\affiliation{
Light-Matter Interactions for Quantum Technologies Unit, Okinawa Institute of Science and Technology Graduate University,
Onna, Okinawa 904-0495, Japan
}

\author{Martina Hentschel}
\affiliation{
Institute of Physics, Technische Universität Chemnitz, D-09107 Chemnitz, Germany
}

\date{September 2025}

\begin{abstract}
Non-Hermitian photonics has attracted significant interest %
and influences several key areas such as optical metamaterials, laser physics, and nonlinear optics. 
While non-Hermitian effects have been widely addressed in two-dimensional systems, we focus on realistic three-dimensional devices. To this end we generalize %
established phase space methods from mesoscopic optics %
and introduce Husimi functions for three-dimensional systems that
deepen the insight and access to the mode morphology and their dynamics. We illustrate that %
four-dimensional Husimi functions can be represented %
using a specific  projection on two dimensions and illustrate it for %
(conical) cylindrical cavities. %
The non-Hermitian character of the intrinsically open 
photonic systems is in particular revealed when 
examining the TE and TM polarization character of the resonance  modes. Unlike the %
2D case, polarization is not conserved in three-dimensional cavities, and 
we use generalized Husimi function to represent the interaction of polarization modes. We find their dynamics to be ruled by 
a network of exceptional points %
in the parameter space spanned by the %
refractive index and the cavity geometry tilt angle. 
This approach not only enhances our understanding of cavity modes but also aids in the design of more efficient photonic devices and systems. 

\vspace{1em}
\noindent
Published in: Physical Review A \textbf{112}, 033528 (2025) \\
DOI: \href{https://doi.org/10.1103/2wrn-z5fg}{10.1103/2wrn-z5fg}
\end{abstract}

\maketitle

\section{Introduction}

Optical microcavities \cite{vahala2003optical, yang2015advances, cao2015dielectric} have proven to be a fascinating and application relevant model system for several decades. 
Their intrinsic openness -- light can enter and leave the cavity by refraction or evanescent coupling -- is the source of fascinating properties and devices ranging from microlasers to sensors.
These open optical microcavities of various geometries have considerably enriched the field of quantum chaos beyond the hard-wall billiard systems typically studied in nonlinear dynamics \cite{stoeckmann1999quantum}. 
Their description 
in both real and phase space has contributed significantly to their understanding in terms of ray-wave correspondence \cite{nockel1997ray, hentschel2002quantum, ketzmerick2022chaotic}, leading to the development and application of deformed microcavity lasers with directional emission \cite{hentschel2009designing, yan2009directional, wang2009deformed, wu2010ultralow, harayama2011two}.
Other notable applications of photonic microsystems include 
cavity QED studies \cite{walther2006cavity, owens2022chiral, taylor2025light}
and single particle detection \cite{baaske2014single, zhi2017single,foreman2017whispering}. 

Often optical cavities can be modeled successfully as two-dimensional (2D) systems, such as 
very long waveguides or flat disc-like cavities. However, experimental systems often have non-trivial geometry in all three dimensions.
In this paper we aim to expand the Husimi function concept to those three-dimensional (3D) optical systems.

Moreover, the openness of such mesoscopic optical cavities yields the remarkable feature of non-Hermitian physics since all resonances are characterized by a complex frequency $\omega$ (with the quality $Q$ factor being related to its imaginary part by  $Q = - \Re(\omega) / 2 \Im(\omega)$) \cite{kato2013perturbation, rotter2009non, ozdemir2019parity} . 
The resulting non-Hermitian effects have been studied in depth in the past years in
a variety of systems %
such as
microlasers \cite{hodaei2014parity, feng2014single, miao2016orbital, peng2016chiral, longhi2018non, zhu2022anomalous},
sensors \cite{wiersig2016sensors, chen2017exceptional, hodaei2017enhanced, wiersig2020review, kononchuk2022exceptional, mao2024exceptional, parto2025enhanced},
encryption \cite{yang2023electromagnetically}, 
topological devices \cite{zhao2019non, liu2019second, weidemann2020topological, xia2021nonlinear, bergholtz2021exceptional} and
optical gyroscopes \cite{sunada2007design, liang2017resonant}.
In addition to optical systems, non-Hermitian properties can also arise in
acoustic \cite{shi2016accessing, gu2022transient}, optomechanical \cite{xu2016topological}, magnetic \cite{yang2018antiferromagnetism} and plasmonic devices \cite{xu2023subwavelength}, as well as integrated electrical circuits \cite{deng2024ultrasensitive}.

The clear distinction between transverse magnetic (TM) and transverse electric (TE) polarization is exploited in the description of 2D systems, where the polarization is fixed. %
This changes in 3D microcavities as
becomes immediately clear when considering a cube resonator: when the polarization-dependent continuity conditions are fulfilled at one pair of side wall interfaces, they cannot be fulfilled for the same polarization for side walls perpendicular to it. Consequently, there will be an interaction of polarization states, though a differentiation of modes according to their (dominant) polarization remains useful. We will illustrate this polarization interaction using generalized Husimi functions and reveal its intimate relation to the non-Hermitian system character.

Phase space methods have a long history in the description of quantum systems \cite{husimi1940some, schleich2011quantum}.
Here we focus on the Husimi function that maps resonance modes in real space into a probability distribution in phase space by computing their overlap with a coherent state (localized at a certain point in phase space)\cite{nonnenmacher1998chaotic,backer2004poincare}.
In contrast to traditional hard-walled billiard systems \cite{crespi1993quantum, nonnenmacher1998chaotic, backer2004poincare, hofferbert2005experimental}, optical microcavities are inherently open and 
both the wave function and its derivative remain nonzero at the system boundary.
At the same time, 
incident and emerging waves (or rays) inside and outside the cavity have to be distinguished. Both issues are connected,
and generalized (2D) Husimi functions associated with incoming and outgoing light can be defined when using both the wave function and its derivative at the boundary \cite{hentschel2003husimi}.
Apart from establishing %
ray-wave correspondence in phase space, Husimi functions have %
been used effectively to investigate for example semiclassical effects such as the Goos-Hänchen shift \cite{lee2005scarred, schomerus2006correcting, unterhinninghofen2008goos}, asymmetric backscattering in cavities \cite{lee2008ray, kullig2016frobenius, rodemund2024coupled}, radiative properties \cite{wiersig2008combining}, and for tayloring the internal dynamics \cite{qian2021regulated, jiang2024coherent,Husimi_Bosch:21} of optical microcavities.

In this paper, we first introduce the concept of mode morphology in a cylindrical waveguide in Section~\ref{sec:mode_structure} and recall 2D Husimi functions.  Next, we develop 3D Husimi functions in Section~\ref{sec:husimi} and, in Section~\ref{sec:cone}, we discuss the interaction of polarization modes using the 3D Husimi functions. We illustrate non-Hermitian effects in the mode dynamics in detail in Section~\ref{sec:nonHerm} and end with a conclusion in Section~\ref{sec:conclusion}.
The data is obtained by performing simulations using COMSOL's frequency domain electromagnetics module.

\section{Cylindrical waveguides and cross-sectional Husimi functions}
\label{sec:mode_structure}

\begin{figure}
    \centering

    \includegraphics[width=1.\columnwidth]{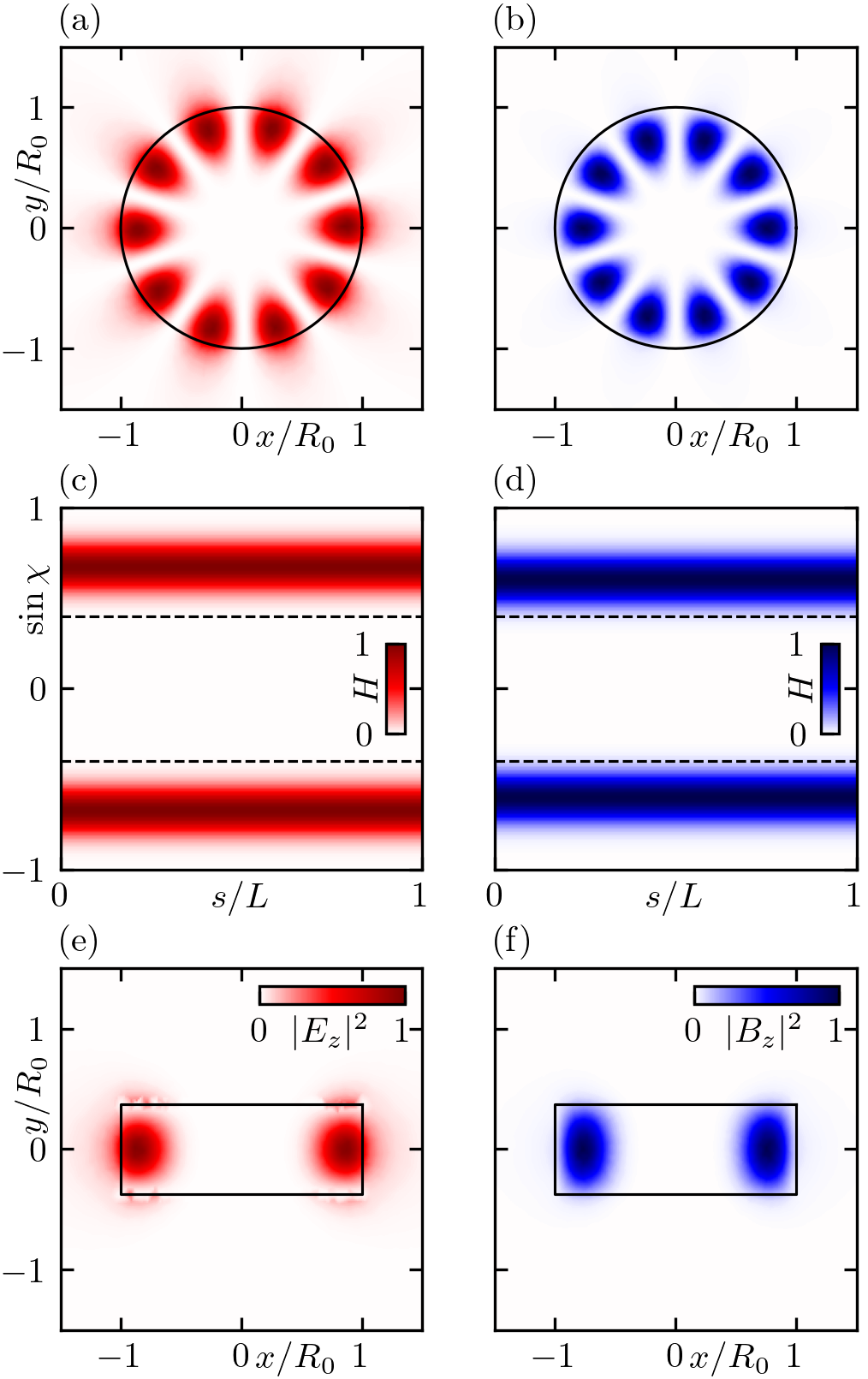}

    \caption{
    (a,b,e,f) Electric (red) and magnetic (blue) field intensities %
    of cylinder modes with quantum numbers $(\qnrr, \qnphi, \qnzz) = (0,5,0)$ and refractive index $n=2.5$.
    The mode in (a,e) is TM-polarized with $\Omega_\mathrm{TM} = 3.361 - \imag \, 0.010$. 
    The mode in (b,f) is TE polarized mode with $\Omega_\mathrm{TE} = 3.578 - \imag  \, 0.014 $.
    (e,f) Incident Husimi functions for the %
    cross-sectional $(x,y)$ plane with (c) $\Omega_\mathrm{TM}^\mathrm{2D} = 2.951 - \imag  \, 0.010$ and (d) $\Omega_\mathrm{TE}^\mathrm{2D} = 3.342 - \imag  \, 0.017$.
    The dashed lines mark minimum angles for total internal reflection.
    }
    \label{fig:mode:flat_cylinder_mode}
\end{figure}

Solving Maxwell's equations for 2D systems, we find that two polarizations can be distinguished. 
For TM (TE) modes, only the out-of plane field component (defining the $z$ direction) $E_z$ ($B_z$) is non-zero, while $\vec{B}$ ($\vec{E}$) possesses in-plane ($x,y$) components only.
The out-of-plane component suffices to characterize the resonance.
It is described by the scalar wave function $\psi(x,y,t) = \psi(x,y) \euler^{- \imag \omega t}$, which needs to satisfy the Helmholtz equation
\begin{equation}\label{eq:2d:helmholtz}
    [\partial_x^2 + \partial_y^2 + \tilde{n}^2(\vec{r}) k^2 ] \, \psi(x,y) = 0
\end{equation}
with $\psi = E_z (H_z)$ for TM (TE) polarized modes and $k = \re(\omega) / \speedoflight$, where $\speedoflight$ is the speed of light in vacuum.
$\tilde{n}(\vec{r})$ is the refractive index function with $\tilde{n}(\vec{r}) = n$ inside and $\tilde{n}(\vec{r}) = 1$ outside the cavity.
For cavities with rotational symmetry around the $z$ axis and radius $R_0$, we can define the normalized frequency $\Omega = \omega R_0 / \speedoflight$. The wave function must satisfy
\begin{equation}
    \left.\psi\right|_{r=R_0} = \mathrm{const} \quad \quad
    \xi \left.\frac{\partial \psi}{\partial r}\right|_{r=R_0} = \mathrm{const}
\end{equation}
with $\xi = 1 \, (1/\tilde{n}^2(\vec{r}))$ for TM (TE) modes at either side of the cavity boundary.
The solution to Eq.~\eqref{eq:2d:helmholtz} reads
\begin{equation}\label{eq:2d:helmholtz_solution}
    \psi(r,\varphi) = \euler^{\pm \imag \qnphi \varphi} \begin{cases}
        c_1 J_\qnphi(n k r ) &, r < R_0 \\
        c_2 H_\qnphi^{(1)}(k r ) &,  r > R_0
    \end{cases}
\end{equation}
with angular momentum quantum number $\qnphi \in \mathbb{Z}$ and the Bessel (Hankel) function $J_m(x)$ ($H_m^{(1)}(x)$) of the first kind.
Note that resonances with $\pm m$ are doubly degenerate.
The values for $c_{1,2}$ are obtained by considering the appropriate boundary conditions.
This introduces a quantization in $r$, which is described by the quantum number $\qnrr \in \mathbb{N}_0$.

Two example solutions are shown in Fig.~\ref{fig:mode:flat_cylinder_mode}(a) and (b).
Fig.~\ref{fig:mode:flat_cylinder_mode}(a) shows the electric field intensity $| \psi|^2$ from Eq.~\eqref{eq:2d:helmholtz_solution} for a TM mode with $(\qnrr,\qnphi) = (0,5)$.
Fig.~\ref{fig:mode:flat_cylinder_mode}(b) shows $| \psi|^2$ for the corresponding TE mode.
These morphologies are also called whispering-gallery modes, as they travel along the cavity surface where they are confined by total internal reflection.
Note that the wave function in Fig.~\ref{fig:mode:flat_cylinder_mode} is calculated using a 3D %
cavity, however the solutions for the out-of-plane component are equivalent to the 2D case \cite{jackson1999classical, bittner2009experimental}.

Phase space methods are a well-developed tool to characterize optical systems with ray and wave dynamics. 
The state of a particle in a two-dimensional billiard system can be described by four generalized coordinates $(\vec{q}, \vec{p})$ with $\vec{q}, \vec{p} \in \mathbb{R}^2$ in a four-dimensional phase space.
In Hamiltonian systems, %
conservation of energy confines a particle %
to a 3D manifold.
In addition, a particle trajectory is completely described by considering its reflection points at the interface boundary, with location  $s$ and angle of incidence $\chi$. %
Consequently, a 2D phase space is sufficient to describe the dynamics of a 2D system.
An example for such a projection is the well-known Poincaré surface of section 
for billiards and ray optics \cite{crespi1993quantum, hentschel2002quantum, hofferbert2005experimental, firmbach2018three}. The Husimi function is the counterpart of the trajectory trace for wave dynamical systems and can be computed based on the wave function and its derivative \cite{hentschel2003husimi, lee2005scarred, lee2008ray, ryu2009coupled, ketzmerick2022chaotic}, see Eq.~\eqref{eq:husimi}
 below.

The 2D Husimi function is based on %
$\psi(s)$ and $\psi'(s)$ where $s$ is the coordinate along the system boundary and $'$ indicates the normal derivative. 
The other phase space coordinate is $\sin \chi$, the sine of the (classical) angle of incidence that is related to angular momentum $m$ by $\sin \chi =  m / n \re (\Omega)$ with $\re (\Omega) = k R_0$.
The Husimi functions $H(s,\sin\chi)$ for the wave function in Figs.~\ref{fig:mode:flat_cylinder_mode}(a,b) are shown in Figs.~\ref{fig:mode:flat_cylinder_mode}(c,d).
The mode is a radially symmetric whispering-gallery mode, and the angular momentum is thus conserved. 
Note that this is the simplest case, as for deformed cavities more intricate wave function (and thus phase space) emerges.

\section{Generalized Husimi functions for three-dimensional cavities}
\label{sec:husimi}

\subsection{Cylindrical cavities of finite height}
\label{sec:husimi:confined_wavefunc}

In dielectric cylinders with height, $h$, the mode is also confined in the $z$-direction.
Thus the $\vec{E}$ and $\vec{B}$ fields can couple and the 3D Helmholtz equation \cite{jackson1999classical}
\begin{equation}\label{eq:3d:helmholtz_vectorial}
    [\Delta + \tilde{n}^2(\vec{r}) k^2] \begin{Bmatrix}
        \vec{E}(\vec{r}) \\ \vec{B}(\vec{r})
    \end{Bmatrix} = \vec{0}
\end{equation}
needs to be solved.
Maxwell's equations also impose the boundary conditions
\begin{gather}\label{eq:3d:boundary_conditions}
\begin{aligned}
    \vec{E}_{1,\parallel} &=\vec{E}_{2,\parallel} &
    n_1^2 \vec{E}_{1,\perp} &= n_2^2 \vec{E}_{2,\perp} \\
    \vec{B}_{1,\parallel} &= \vec{B}_{2,\parallel} &
    \vec{B}_{1,\perp} &= \vec{B}_{2,\perp} \quad
\end{aligned}
\end{gather}
with parallel ($\parallel$) and perpendicular ($\perp$) field components relative to the cavity interface
and subscripts 1 (2) referring to points inside (outside) the cavity.

Applying the appropriate boundary conditions leads to a quantization in the $z$-direction,  described by the quantum number $\qnzz \in \mathbb{N}_0$.
So in general, modes are identified by their polarization and the three quantum numbers $(\qnrr, \qnphi, \qnzz)$.
Examples for confinement in the $z$ direction for TM and TE modes are shown in Fig.~\ref{fig:mode:flat_cylinder_mode}(e) and (f).
The TM mode is more confined in $z$ direction due to the discontinuity in $E_z$ described in Eq.~\eqref{eq:3d:boundary_conditions}.

To examine the interaction between electric and magnetic field components we use the ansatz $\vec{E}(\vec{r}, t) = \vec{E}(x,y) \euler^{\imag k_z z - \imag \omega t}$ for Eq.~\eqref{eq:3d:helmholtz_vectorial},
and separate the transversal component $\vec{E}_\mathrm{t}$ with
\begin{equation}
    \vec{E}_t = (\vec{e}_z \times \vec{E}) \times \vec{e}_z \: .
\end{equation}
The expression for $\vec{B}$ is similar.
Applying Maxwell's equations gives
\begin{equation}\label{eq:3d:coupling}
\begin{split}
    \vec{E}_\mathrm{t} & = \frac{\imag}{k_{xy}^2} \,  (k_z \nabla_\mathrm{t} E_z - k \speedoflight \, \vec{e}_z \times \nabla_t B_z) \quad \text{and}\\
    \vec{B}_\mathrm{t} & = \frac{\imag}{k_{xy}^2} \,  (k_z \nabla_\mathrm{t} B_z + \frac{n^2 k}{\speedoflight} \, \vec{e}_z \times \nabla_\mathrm{t} E_z) \quad ,
\end{split}
\end{equation}
for the coupling between %
the in- and out-of-plane field components, with $n^2k^2 = k_{xy}^2 + k_z^2$ and transversal nabla operator $\nabla_\mathrm{t} = \nabla - \vec{e}_z \partial_z$.

\subsection{Extension of the Husimi formalism to three-dimensional cavities}
\label{sec:husimi:extension}

For realistic 3D systems that we now describe,
the new generalized coordinates $(\vec{q}, \vec{p})$ with $\vec{q}, \vec{p} \in \mathbb{R}^3$ result in a 6D phase space. 
It reduces to four dimensions after considering conservation of energy and projection to a surface of section that we choose to be the cylinder mantle here. %
We map the wave function onto the cylinder mantle using $\psi(\varphi,z) \rightarrow \psi(s, z)$, where $s= \varphi R_0$ is the arclength along the cavity surface.
The resulting boundary conditions are periodic  in $s$ with $\psi(s,z) = \psi(s \pm 2 \pi R_0,z)$.
Here, we focus on the mantle as the main surface contribution, although top and bottom surfaces could be included straightforwardly when needed, e.g., for flat cavities.

\begin{figure}
	\centering
	\includegraphics[width=0.7\columnwidth]{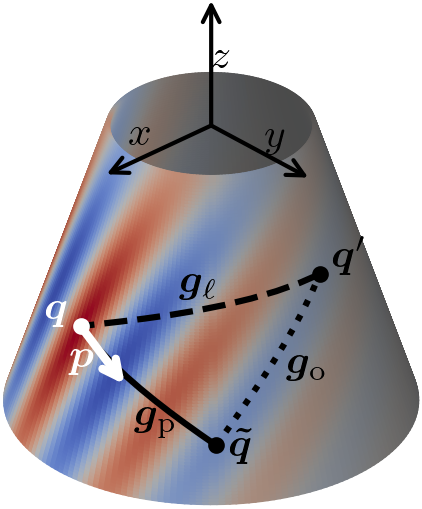}
	\caption{
		Coherent state $\xi(\vec{q}'; \vec{q}, \vec{p})$ on the mantle of a slanted, cylindrical cavity. 
		It is measured at $\vec{q}'$, where the value is determined by the geodesics $\geodesiclength,\geodesicphase,\geodesicortho$ and the point $\vec{\tilde{q}}$.
	}
	\label{fig:husimi:surfacestate}
\end{figure}

In the following we will focus on the inner incident Husimi function, as we are interested in the dynamics inside the cavity.
We consider the overlap of the wave function on the boundary with a two-dimensional coherent state
\begin{multline} \label{eq:husimi:cohstate}
    \xi(\vec{q}'; \vec{q}, \vec{p}) = \frac{1}{\sqrt{\pi \sigma^2}} \times \\
    \exp \left[
    - 
    \distancefunc^2(\vec{q}'; \vec{q})
    / 2 \sigma^2
    - \imag \, 
    \phasefunc(\vec{q}'; \vec{q}, \vec{p})
    \right]
\end{multline}
with 
position $\vec{q}$, momentum %
$|\vec{p}|\leq nk$ on the cavity surface, and a Gaussian spatial profile characterized by $\sigma^2$.
$\distancefunc$ is the geodetic distance %
between $\vec{q}$ and $\vec{q'}$ along $\geodesiclength$ (see Fig.~\ref{fig:husimi:surfacestate}). 
The second term is related to the coherent state's angular momentum that is encoded in the phase $\phasefunc$ %
of the coherent state.
The phase at $\vec{q}'$ is obtained by projection on $\geodesicphase$ via a geodesic $\geodesicortho$ orthogonal to $\geodesicphase$.
We call the projected point $\vec{\tilde{q}}$, where the phase is given as $\phasefunc(\vec{\tilde{q}};\vec{q}, \vec{p}) = |\vec{p}| \, \distancefunc(\vec{\tilde{q}};\vec{q}) = \phasefunc(\vec{q}';\vec{q}, \vec{p})$ that reduces to the well-known scalar product form of the phase term, $\vec{p} \, \vec{r_{\vec{q},\vec{q'}}}$ (with
$\vec{r_{\vec{q},\vec{q'}}}$ the vector from $\vec{q}$ to $\vec{q'}$) in euclidean geometries.

We then define the  overlap of the coherent state with the wave function $\psi$ at the boundary as
\begin{subequations}
\begin{align}
    h(\vec{q}, \vec{p}) &= \int \psi(\vec{q}') \, \xi(\vec{q}'; \vec{q}, \vec{p}) \, \mathrm{d}^2\vec{q}' \quad \text{and} \\
    h'(\vec{q}, \vec{p}) &= \int \psi'(\vec{q}') \, \xi(\vec{q}'; \vec{q}, \vec{p}) \, \mathrm{d}^2\vec{q}' \quad ,
\end{align}
\end{subequations}
where $\psi'$ is the radial derivative inside the cavity.
Finally, we arrive at the expression for the Husimi function
\begin{multline}
    H(\vec{q}, \vec{p}) = \\
    \frac{n k}{2 \pi} \left|- \mathcal{F}(p) \, h(\vec{q}, \vec{p}) + \frac{i}{k \mathcal{F}(p)} \, h'(\vec{q}, \vec{p}) \right|^2
    \label{eq:husimi}
\end{multline}
with angular momentum weight $\mathcal{F}(p) = \sqrt{n} (1-\sin^2\chi_0)^{1/4}$.

\begin{figure}
	\centering
	
	\includegraphics[width=.6\columnwidth]{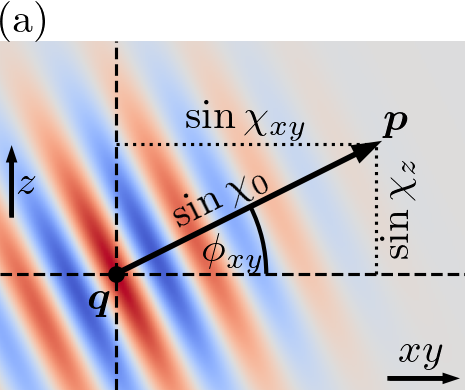}
	
	\vspace{.5em}
	
	\includegraphics[width=1.\columnwidth]{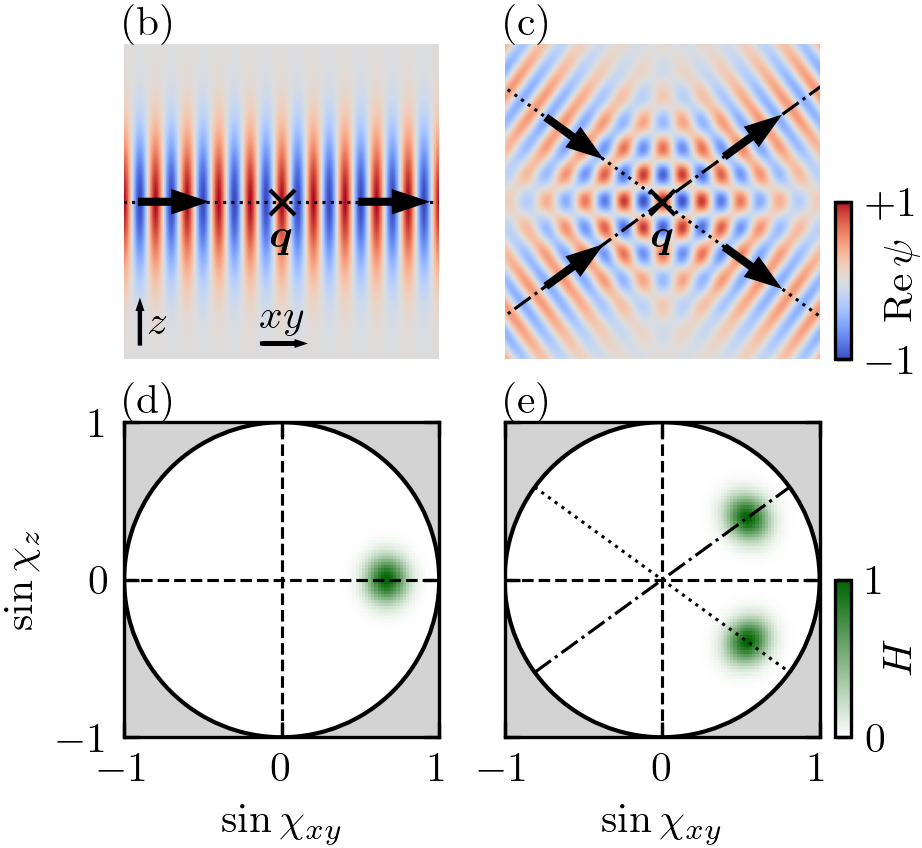}
	\caption{
		(a) Coherent state coordinates,
		(b,c) example wave patterns on the cavity surface and (d,e) their Husimi function signatures taken at $\vec{q}$.
	}
	\label{fig:husimi:mode_morphology_dependence}
\end{figure}

For cylindrical (or conic) cavities the geodesics can be obtained trivially, as all straight lines on the unrolled surface are geodesics.
Then, $\distancefunc$ and $\phasefunc$ can be evaluated as if the cavity mantle were a flat surface using scalar products.
For the phase space coordinates we follow the conventional notation of the Husimi function formalism \cite{hentschel2003husimi}.
For cylindrical 3D cavities the Husimi function
 $   H = H(\vec{q}, \vec{p}) $ %
depends on the spatial coordinates $\vec{q} = (s,z)$ and on the momentum coordinates $\vec{p} = (n k \sin \chi_{xy}, n k \sin \chi_z)$, see Fig.~\ref{fig:husimi:mode_morphology_dependence}(a), distinguishing between components orthogonal and parallel to the $z$-axis.
The momentum coordinates can also be written in polar representation $(n k\sin \chi_0, \phi_{xy})$ with total momentum in the surface (mantle) plane $n^2 k^2\sin^2 \chi_0 = |\vec{p}|^2 = n^2 k^2 (\sin^2 \chi_{xy} + \sin^2 \chi_{z})$ and angle $\phi_{xy}$ to the $xy$-plane where $\tan \phi_{xy} = \sin \chi_{z} / \sin \chi_{xy}$.

To illustrate this generalized Husimi function in the new phase space spanned by $\sin \chi_{xy}$ and $\sin \chi_z$, it is helpful to consider a simple test system:
A wave function $\psi$ is given by a linear combination of plane waves, which propagate along a two-dimensional surface as shown in Fig.~\ref{fig:husimi:mode_morphology_dependence}(b-c).
The Husimi function is calculated for a representative, fixed position $\vec{q}$ while the momentum $\vec{p}$ is varied.
A single plane wave can be identified trivially in real space and gives a single peak in phase space in Fig.~\ref{fig:husimi:mode_morphology_dependence}(d).
For two (or more) overlapping plane waves in Fig.~\ref{fig:husimi:mode_morphology_dependence}(c) the field function can become arbitrarily complex, making it impossible to distinguish components in real space.
The phase space recovers the constituent plane waves and their parameters as shown in Fig.~\ref{fig:husimi:mode_morphology_dependence}(e). 

In %
2D systems, the mode is captured %
by the out-of-plane (or $z$-) component, resulting in the natural choice of $\psi = E_z (H_z)$ for TM (TE) polarization as discussed above.
In %
3D systems all six field components are nonzero, resulting in a degree of freedom concerning the choice of $\psi$.
In cylindrical systems, the field components are related via Eq.~\eqref{eq:3d:coupling}, resulting in qualitatively similar field distributions for all components.
The quantitative difference between field components lies in the number of oscillations in $z$ for a mode with a given $\qnzz$.
The $E_z$ component of a TM mode has $\qnzz+1$ intensity maxima in the $z$ direction, while all other components have $\qnzz+2$ intensity maxima.

In the following, we utilize the field components parallel to the cavity mantle surface for phase space calculations. This is the most natural choice as $E_z$ and $B_z$ are continuous on the cylinder mantle, and they are also the components of choice in the 2D case.
The conventional Husimi function for 2D cavities is recovered for $h/R_0 \rightarrow \infty$ (or $k_z \rightarrow 0$) when the values are taken along $\sin \chi_z = 0$ and $\vec{q}$ is moving along a line in the symmetry plane.

\subsection{Husimi functions for cylindrical cavities}
\label{sec:husimi:application}

\begin{figure}
    \centering

    \begin{subfigure}{.7\columnwidth}
    \centering
        \includegraphics[width=.8\textwidth]{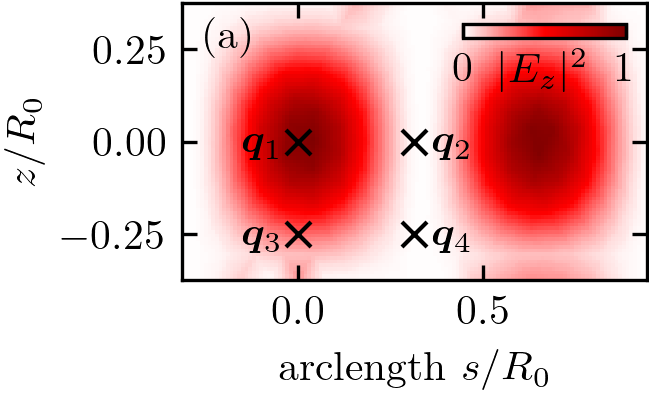}

    \end{subfigure}

    \vspace{.5em}
    
    \begin{subfigure}{.8\columnwidth}
    \centering 
        \includegraphics[width=.98\textwidth]{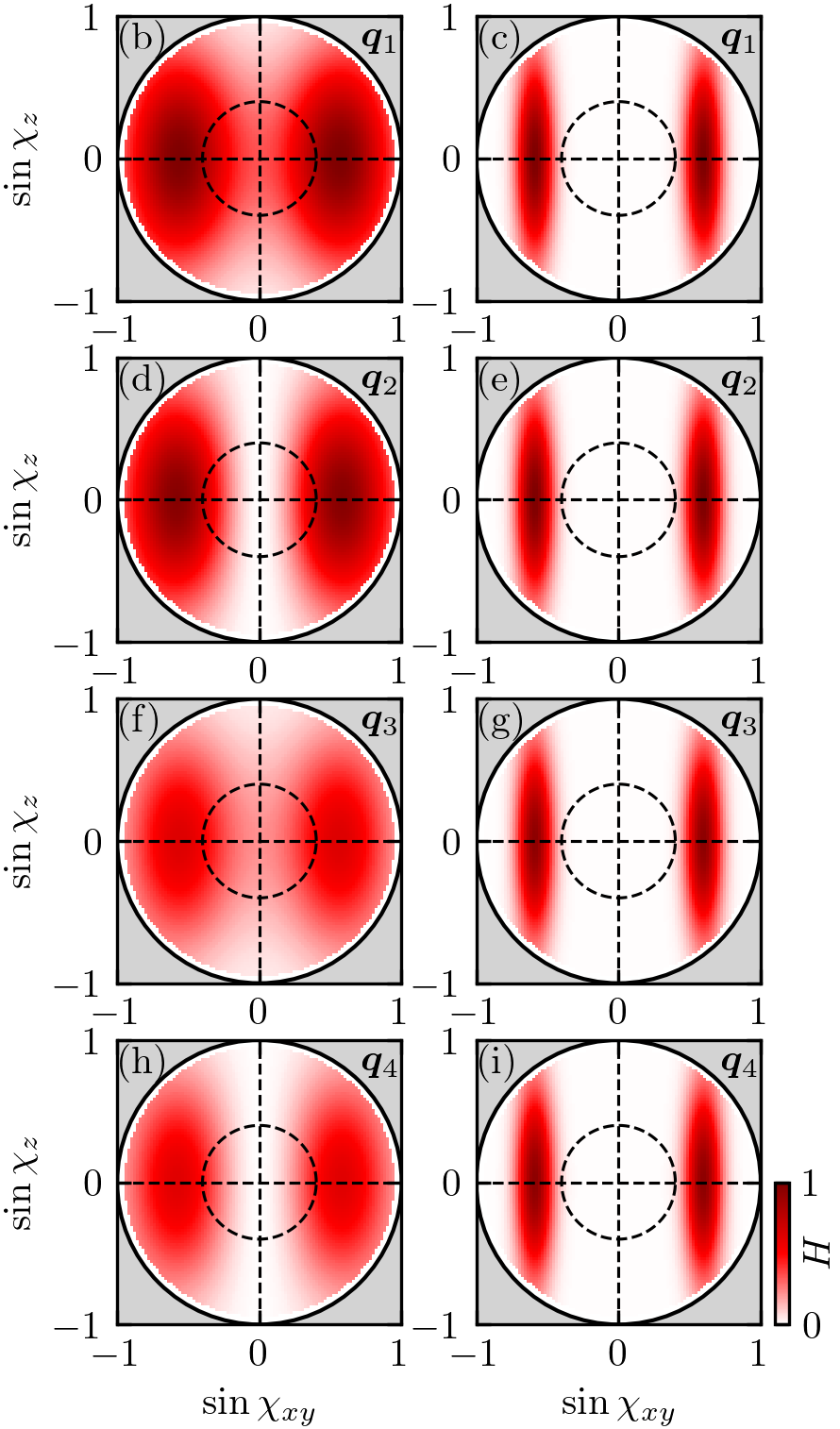}

    \end{subfigure}

    \caption{
    Husimi functions for the TM mode with $(\qnrr, \qnphi, \qnzz)=(0,5,0)$ in a dielectric cylinder with $H/R_0 = 0.75$ for various positions on the mantle.
    (a) The chosen coherent state positions, $\vec{q}_i$.
    The components of $\vec{q}_i$ can assume values of $q_s/R_0 = 0.00, 0.31$ and $q_z/R_0 = 0.00, -0.25$.
    The results are shown with coherent state sizes of (b,d,f,h) $\sigma/R_0 = 0.32$ (or $\mathrm{FWHM} = \lambda / n$) and (c,e,g,i) $\sigma/R_0 = 0.95$ (or $\mathrm{FWHM} = 3 \lambda / n$).
    }
    \label{fig:husimi:vary_cohpos_cohsize}
\end{figure}

\begin{figure}
    \centering

    \begin{subfigure}{0.32\columnwidth}
        \includegraphics[width=1.\textwidth]{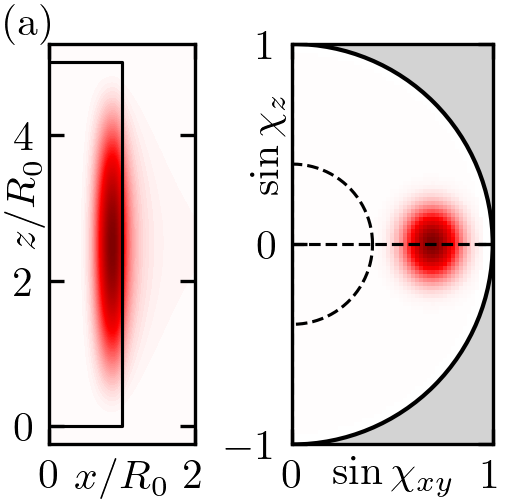}      
    \end{subfigure}%
    \begin{subfigure}{0.32\columnwidth}
        \includegraphics[width=1.\textwidth]{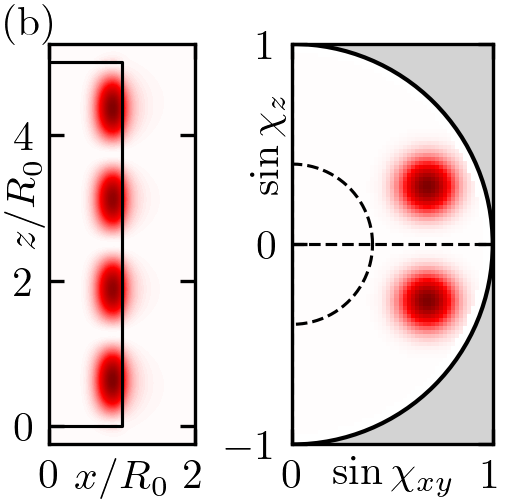}
    \end{subfigure}
      \begin{subfigure}{0.32\columnwidth}
        \includegraphics[width=1.\textwidth]{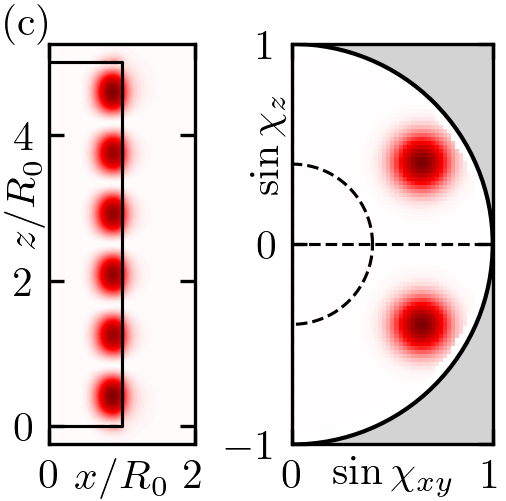}   
    \end{subfigure}

    \caption{
    Field intensity (left) and Husimi functions (right) for  TM modes, with $(\qnrr,\qnphi)=(0,6)$ and various $\qnzz$. 
    The $\qnzz$ and frequencies are (a) $(\qnzz,\Omega) = (0, 3.4402 - \imag \, 0.0049)$, (b) $(\qnzz,\Omega) = (3, 3.5578 - \imag \, 0.0014)$ and (c) $(\qnzz,\Omega) = (5, 3.7182 - \imag \, 0.0006)$.
    The phase space functions are shown for $\sin \chi_{xy} > 0$, the rest is given by symmetry.
    }
    \label{fig:husimi:vary_nz}
\end{figure}

The Husimi function for the $(\qnrr, \qnphi, \qnzz)=(0,5,0)$-TM and TE modes with $h/R = 0.75$ are shown in Fig.~\ref{fig:mode:flat_cylinder_mode}(e) and (f).
The phase space figures are calculated by using a constant coherent state position $\vec{q} = \mathrm{const}$ and varying the momentum coordinates $\vec{p}$.
This gives a circular border for the parameter space from the constraint $\sin^2 \chi_{xy} + \sin^2 \chi_{z} < 1$.
Due to the rotational symmetry, only coherent state positions with angles $\varphi \in [0, \pi / 2 m]$ need to be considered.
Given the mirror symmetry at $z=0$, we focus on coherent states with $z \in [0,-h/2]$ without loss of generality.

A sharp peak is present at $\sin \chi_{xy} = \pm \qnphi / n \, \re \Omega$ for all values of $\sin \chi_z$.  This is analogous to the result of the 2D cavity in Figs.~\ref{fig:mode:flat_cylinder_mode}(c,f).
The phase space is mirrored along $\sin \chi_{xy} = 0$ because the standing wave considered here is a superposition of two traveling waves with $\pm \qnphi$.
In addition, the existence of the geometry symmetry plane, $z=0$, results in a phase space mirror symmetry  with axis $\sin \chi_z = 0$. 
For the TE mode, the intensity maximum is closer to the origin which reflects its shorter lifetime since a larger part of the phase space intensity is located in the leaky region.

Another degree of freedom we need to take into account is the localization of the coherent state, given by its standard deviation $\sigma$ in Eq.~\eqref{eq:husimi:cohstate}.
Fig.~\ref{fig:husimi:vary_cohpos_cohsize} compares the phase spaces for two values of $\sigma$ for various points $\vec{q}_i$ on the cylinder surface.
Figures~\ref{fig:husimi:vary_cohpos_cohsize}(b,d,f) and (h) show the results for strong localization with $\sigma/R_0 = 0.32$,  corresponding to a full-width at half-maximum of $\lambda/n$, where $\lambda$ is the vacuum wavelength.
The intensity distribution is different between $\vec{q}_1$ and $\vec{q}_2$ in Figs.~\ref{fig:husimi:vary_cohpos_cohsize}(b) and (d).
$\vec{q}_1$ is located on an intensity maximum, leading to $H(\vec{q}=\vec{q}_1, \vec{p}=\vec{0}) > 0$.
The nodal line at $\vec{q}_2$ results in $H(\vec{q}=\vec{q}_2, \vec{p}=\vec{0}) = 0$, since the fields of two neighboring intensity maxima average to zero due to their different signs.
The Husimi functions in Figs.~\ref{fig:husimi:vary_cohpos_cohsize}(f,h) are taken at points closer to the cavity boundary.
The intensity at $\vec{q}=\vec{q}_3, \vec{q}_4$ is significantly lower than at $\vec{q}=\vec{q}_1, \vec{q}_2$, which is due to the lower overall field intensity at the cylinder end faces.
Spatial dependence is very pronounced for localized coherent states and is %
lost for the %
weakly localized state with $\sigma = 0.95$, see Figs.~\ref{fig:husimi:vary_cohpos_cohsize}(c,e,g) and i.

Fig.~\ref{fig:husimi:vary_nz} explores the dependence of the Husimi function on the quantization in $z$ for a cylinder with $h/R = 5$.
It shows the behavior of the Husimi function with $\sigma/R_0 = 0.95$ for $\qnzz = 0,3,5$ for TM modes with $(\qnrr,\qnphi)=(0,6)$.
The wave functions are now described by sharply localized intensity maxima, in contrast to the previously spread out maxima in Fig.~\ref{fig:husimi:vary_cohpos_cohsize}.
When $\qnzz$ is increased in Figs.~\ref{fig:husimi:vary_nz}(b) and (c), the angle $\varphi_{xy}$ of the maximum increases.
This is related to the distribution of wavenumber components in the in- and out-of-plane components.
The mode performs whispering-gallery type motion in $xy$ and a "bouncing ball" type motion in $z$.
The rays can be thought of as angled at $\tan \phi_{xy} = k_z / k_{xy}$ (compare Fig.~\ref{fig:husimi:mode_morphology_dependence}(a)).
Increasing $\qnzz$ results in more energy being stored in the bouncing ball motion, thereby increasing $\sin \chi_z$ of the intensity maximum.
The symmetry in the $xy$-plane results in two symmetry-related signatures at $\pm \sin \chi_z$.
The wave function on the cylinder mantle displays a "checkerboard" pattern, which can be recovered by the superposition of two plane waves at angles $\pm \varphi_{xy}$ (compare Figs.~\ref{fig:husimi:mode_morphology_dependence}(c,e)).
For TE modes the behavior is similar,
which is demonstrated in Appendix~\ref{sec:appendix:husimi_vary_nz}.

\label{sec:cone}

\subsection{Wave and Husimi function morphology evolution in conical cylinders}
\label{sec:cone:morphology}

\begin{figure}[!h]
    \begin{subfigure}[c]{.50\columnwidth}
        \includegraphics[width=1.\textwidth]{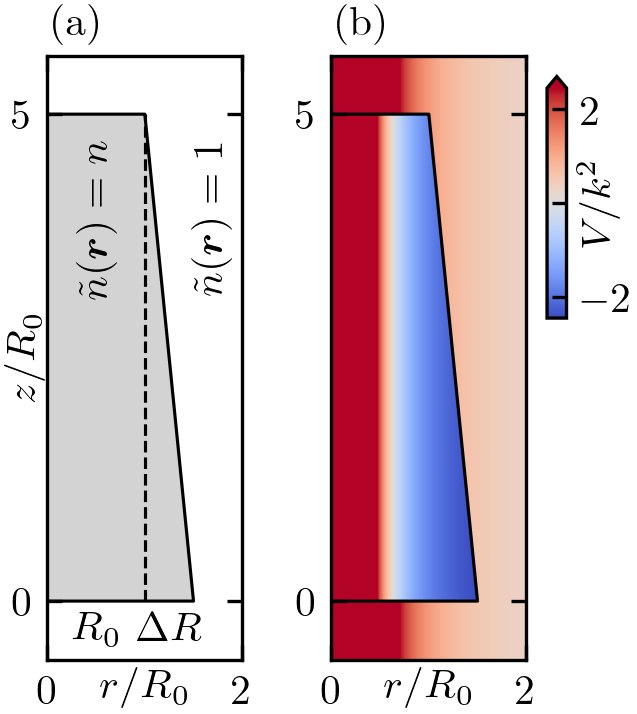}
    \end{subfigure}

    \vspace{.7em}

    \begin{tblr}{c|[dashed]c|[dashed]c}
         & $\cyldeform = 0$ & $\cyldeform=0.1$ \\
        \hline[dashed]
        \rotatebox[origin=c]{90}{$\qnzz = 0$}
        & 
        \begin{minipage}{.38\columnwidth}
            \centering
            \includegraphics[width=1.\textwidth]{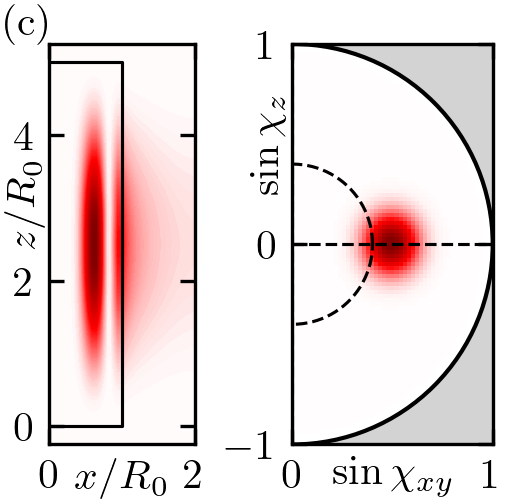}

            $\Omega = 4.820 - \imag \, 0.038$
        \end{minipage}
         & 
        \begin{minipage}{.38\columnwidth}
            \centering
            \includegraphics[width=1.\textwidth]{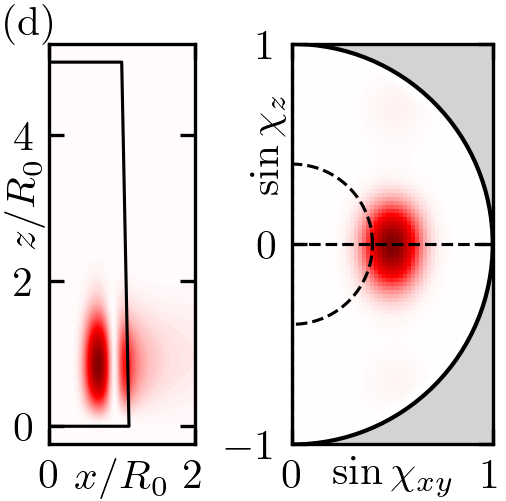}

            $\Omega = 4.487 - \imag \, 0.032$
        \end{minipage} \\
        \hline[dashed]
        \rotatebox[origin=c]{90}{$\qnzz = 2$}
        & 
        \begin{minipage}{.38\columnwidth}
            \centering
            \includegraphics[width=1.\textwidth]{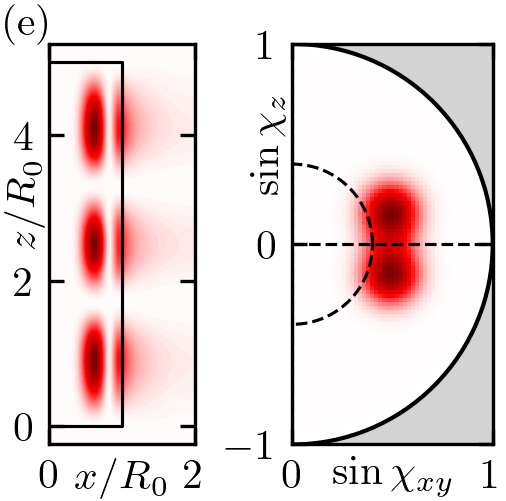}

            $\Omega = 4.854 - \imag \, 0.035$
        \end{minipage} & 
        \begin{minipage}{.38\columnwidth}
            \centering
            \includegraphics[width=1.\textwidth]{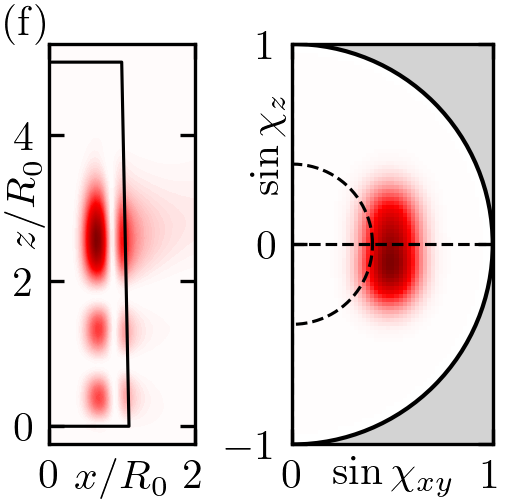}

            $\Omega = 4.630 - \imag \, 0.028$
        \end{minipage} \\
        \hline[dashed]
        \rotatebox[origin=c]{90}{$\qnzz = 7$}
        & 
        \begin{minipage}{.38\columnwidth}
            \centering
            \includegraphics[width=1.\textwidth]{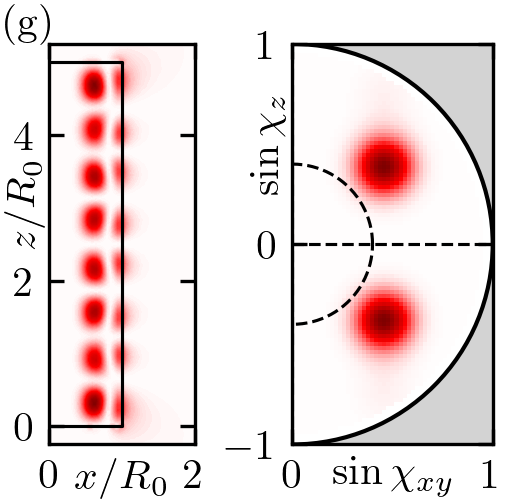}

            $\Omega = 5.185 - \imag \, 0.013$
        \end{minipage} & 
        \begin{minipage}{.38\columnwidth}
            \centering
            \includegraphics[width=1.\textwidth]{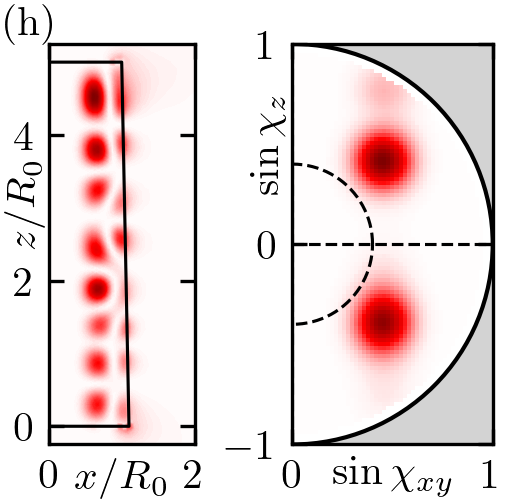}

            $\Omega = 5.002 - \imag \, 0.015$
        \end{minipage} 
    \end{tblr}

    \centering
    \caption{
    (a) System geometry and (b) optical potential of the slanted cone.
    (c-h) $E_\parallel$ field intensity (left) and Husimi functions at $z=0$ (right) of TM modes with (c,d) $\qnzz = 0$, (e,f) $\qnzz = 3$ and (g,h) $\qnzz = 7$ 
    for deformation parameters (c,e,g) $\cyldeform = 0$ and (d,f,h) $\cyldeform = 0.1$.
    }
    \label{fig:cone:mode_morphology}
\end{figure}

For dielectric cylinders the $\vec{E}$ and $\vec{B}$ fields are coupled (see Eq.~\eqref{eq:3d:coupling}).
To control the coupling strength we now introduce a slight deformation to the cylinder.
The radius of the cylinder end face is increased by a perturbation, $\Delta R$. 
We assume a linear dependence of the radius along $z$, such that
\begin{equation}\label{eq:cone:radius}
    R(z)/R_0 = 1 + \cyldeform (1 - z / h) \:,
\end{equation}
with deformation strength $\cyldeform = \Delta R / R_0$ and $z \in [0,h]$.
In the following, we focus on systems with $h/R_0=5$ and modes with azimuthal quantum number $\qnphi=6$.
This geometry can also be described as a cone stump with opening angle $\alpha$ where $\tan \alpha = \Delta R / h = \cyldeform/5$, see  Fig.~\ref{fig:cone:mode_morphology}(a).
This geometry is relevant for experimental setups, as microcavity side walls are usually slightly slanted due to fabrication methods.
Fig.~\ref{fig:cone:mode_morphology}(b) shows the resulting effective optical potential
\begin{equation} \label{eq:cone:optical_potential}
    V(\vec{r}) = [1-\tilde{n}^2(\vec{r})] k^2 + \frac{m^2}{r^2}
\end{equation}
of a slanted cone.
The potential has its minimum at the cylinder bottom, suggesting that fundamental modes will be located at $z\approx 0$.

To explore the influence of the deformation, $\cyldeform$, on any given mode, we investigate the evolution of the resonance morphology.
To this end we focus on the field components which are parallel  to the sidewalls $E_\parallel$ ($B_\parallel$) for TM (TE) polarized modes, as they are continuous at the interface.
The electric field  is defined as
\begin{equation}
    E_\parallel(\vec{r}) = E_z(\vec{r}) \cos \alpha - E_r(\vec{r}) \sin \alpha, \quad 
\end{equation}
with a similar expression for $B_\parallel$.

The left panels in Fig.~\ref{fig:cone:mode_morphology}(c-h) show the field functions of three different TM modes with varying excitations in the $z$-direction.
The real parts of the eigenfrequencies decrease for all modes. 
This is due to $n \re(\Omega) R_\mathrm{eff}/R_0 \approx \mathrm{const}$ for any given mode.
The deformation $\cyldeform$ increases the effective radius, with $R_0 (1+\cyldeform) \propto R_\mathrm{eff}$, 
which, in turn, implies $1+\cyldeform \propto \re(\Omega)^{-1}$.
The field intensity shifts to the bottom of the cavity for low $\qnzz$ (see Figs.~\ref{fig:cone:mode_morphology}(c,d) and (e,f)) due to the lower optical potential according to Eq.~\eqref{eq:cone:optical_potential}.
For modes with high $\qnzz$, i.e. Figs.~\ref{fig:cone:mode_morphology}(g,h), the morphological adaptability is constrained by the necessity to remain orthogonal to the modes with lower $\qnzz$. 
The energetically lower states already occupy the space at the bottom  
where of the radius is largest, making it  less accessible %
for the higher states.

In Figs.~\ref{fig:cone:mode_morphology}(f,h), we see that the field concentrates in the top intensity maximum when $\cyldeform$ is increased; this is also a byproduct of the orthogonality requirement.
Hence, less field intensity leaks out of the cavity top face for modes with low $\qnzz$, as the field intensity is very low in that area.
The imaginary component of $\Omega$ increases accordingly.
In contrast, modes with high $\qnzz$ exhibit decreasing imaginary components.
The proximity of the uppermost intensity maximum to the cylinder top face results in 

increased losses, induced by the sharp corner.
Similar behavior is also prevalent in 2D cavities, where sharp edges are major sources of refractive escape \cite{braun2000hexagonal, wiersig2006formation}.
With this mechanism, the deformation $\cyldeform$ can in- or decrease the resonance lifetime, depending on the excitation $\qnzz$ in $z$.

\begin{figure*}
    \centering

    \begin{subfigure}[c]{.83\textwidth}
        \includegraphics[width=1.\textwidth]{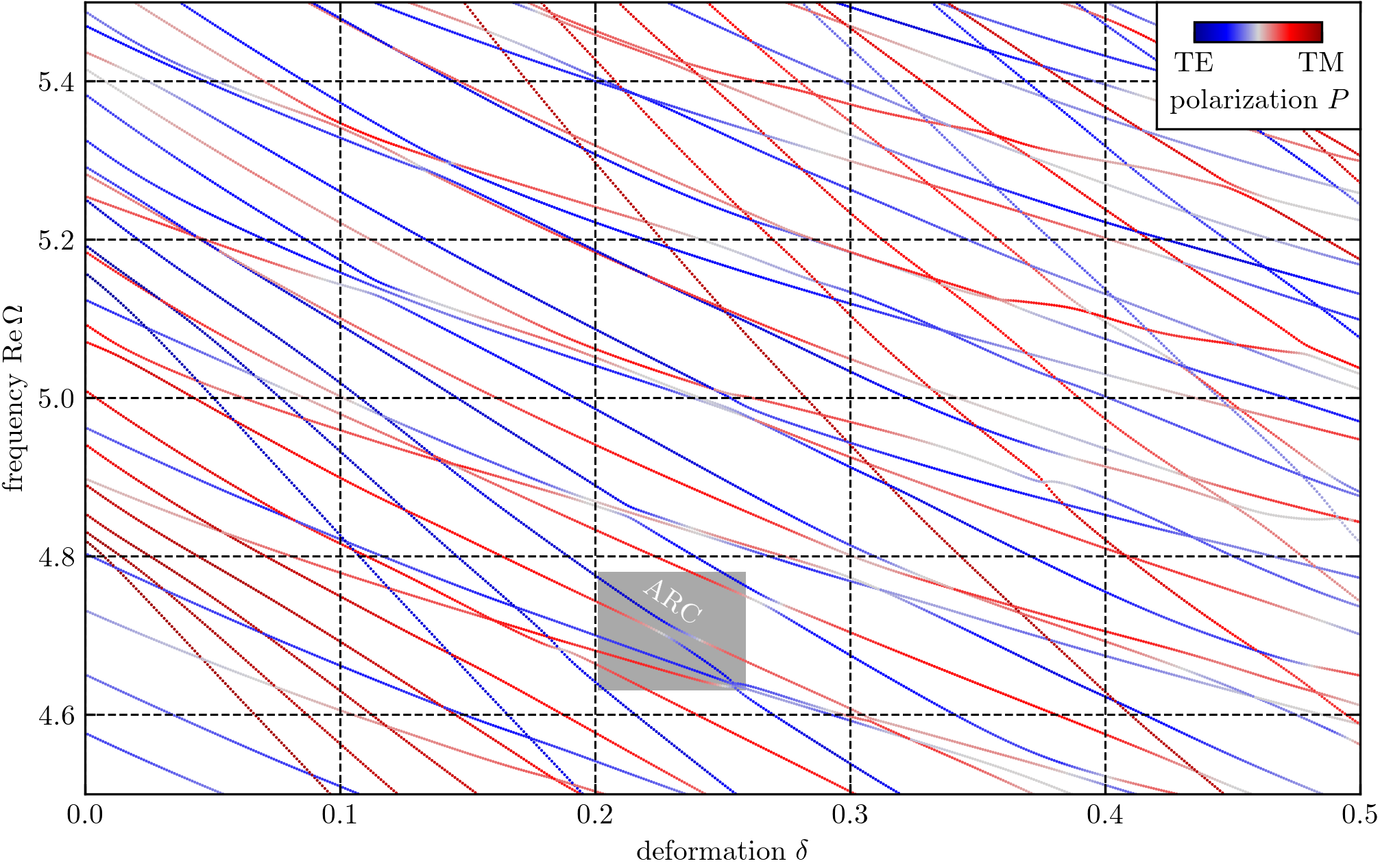}
    \end{subfigure}

    \caption{
    Real parts of eigenvalues for the slanted cone.
    All modes with $\re \Omega \in [4.5,5.5]$ are shown.
    Color denotes polarization $P \in [-1,1]$ according to Eq.~\eqref{eq:cone:def_polarization}, with red (blue) corresponding to TM (TE) polarized modes.
    Grey dots show unpolarized modes with $P\approx 0$.
    The spectrum at $\cyldeform = 0$ reflects the undeformed cylinder.
    The dark grey area marks the location of the avoided resonance crossing in Fig.~\ref{fig:cone:arc_riemann}.
    }
    \label{fig:cone:spectrum}
\end{figure*}

Husimi functions at $z=h/2$ are shown in the right hand panels of Figs.~\ref{fig:cone:mode_morphology}(c-h).  
Here, the wave function on the cavity surface was projected onto a Euclidean plane, as discussed in Sec.~\ref{sec:husimi:extension}.
Similarly to the cylinder, the cone surface can be unrolled using an appropriate transformation, $\psi(\varphi,z) \rightarrow \psi(s,\tilde{z})$, where $\tilde{z}$ is the coordinate in the unrolled plane.
The details can be found in Appendix~\ref{sec:appendix:cone_mapping}.

For $\cyldeform=0$ in Figs.~\ref{fig:cone:mode_morphology}(c,e,g), the behavior corresponds to $\qnzz$ as discussed above.
Due to the mirror symmetry at $z=h/2$, the phase space is symmetric with respect to $\sin \chi_z = 0$.
Generally, this changes for nonzero $\cyldeform$.
The broken symmetry is visible in Figs.~\ref{fig:cone:mode_morphology}(d,f,h), where the Husimi functions for $\cyldeform=0.1$ are shown.
The signature of the mode with $\qnzz=0$ in Fig.~\ref{fig:cone:mode_morphology}(d) is only slightly distorted in comparison to Fig.~\ref{fig:cone:mode_morphology}(c).
Even with the change in morphology, the predominant motion is still along the cavity boundary, resulting in only the signature at $\sin \chi_z = 0$.
In Fig.~\ref{fig:cone:mode_morphology}(f) the asymmetry is more pronounced for the mode with $\qnzz = 2$.
The uppermost field-intensity maximum is dominant in size and intensity due to the deformation.
Due to this spatial expansion, the two Husimi maxima merge into one, which is shifted into the area of $\sin \chi_z < 0$.
The overall intensity inside the cavity at $z>h/2$ is drastically lower than for $z<h/2$ due to the mode shift into areas of higher radius.
For high $\qnzz$, e.g. $\qnzz = 7$ in Fig.~\ref{fig:cone:mode_morphology}(h), the phase space distribution is more balanced, which is a result of the more evenly distributed field function.
However, a distortion of the phase space in comparison to Fig.~\ref{fig:cone:mode_morphology}(g) is visible, emphasizing that the mode structure is more complex for the deformed cavity.
It is evident that the Husimi function encodes the quantum numbers and the local morphology of a mode.

Note that the Husimi function depends on the position $\vec{q}$ of the coherent state.
This is further explored in 
Appendix~\ref{sec:appendix:husimi_spatial_dependence}.

\section{Non-Hermitian mode dynamics}
\label{sec:nonHerm}

\subsection{TE-TM mode coupling and formation of exceptional points}
\label{sec:cone:polarization_coupling}

For small deformations $\cyldeform$ the TM and TE polarizations are, in principle, still distinguishable by comparing the in- and out-of-plane components of the electric and magnetic fields. 
Due to the slanted sidewalls, more intricate mode patterns emerge that do not clearly belong to either category. 
To characterize the polarization of these hybrid states, we introduce the polarization parameter $P$, with $P \in [-1,1]$, which describes the alignment of the two fields,
\begin{equation}\label{eq:cone:def_polarization}
    P = \frac{\tilde{E} - \tilde{B}}{\tilde{E} + \tilde{B}},
\end{equation}
with relative field strengths
\begin{subequations}
\begin{align}
    \tilde{E} &= \max_{\vec{r}}(|E_z(\vec{r})|^2) / \max_{j, \vec{r}} (|E_j(\vec{r})|^2) \quad  \\
	\tilde{B} &= \max_{\vec{r}}(|B_z(\vec{r})|^2) / \max_{j, \vec{r}} (|B_j(\vec{r})|^2), \quad 
\end{align}
\end{subequations}
where the subscript $i$ corresponds to the three field components with $j \in \{r, \varphi, z \}$.
The expression $\max_{\vec{r}}$ takes the spatial maximum, 
while $\max_{j,\vec{r}}$ additionally takes the maximum over all three field components.
It follows that TM (TE) modes yield values of $P \approx 1 (-1)$.
In the TM case, the $E_z$ component is the strongest, resulting in $\tilde{E} \approx 1$. 
The in-plane components of the magnetic field are dominating intensity-wise, leading to $\tilde{B} \approx 0$ and thus $P 
\approx 1$. 
The opposite is true for TE modes.

The eigenvalue spectrum of the weakly deformed cylinder is shown in Fig.~\ref{fig:cone:spectrum}.
For each deformation, $\cyldeform$, the eigenmodes are assigned a color according to their polarization, $P$,  calculated from Eq.~\eqref{eq:cone:def_polarization}.
Red (blue) dots correspond to TM (TE) modes.
All quasibound states with $\re \Omega \in \left[4.5, 5.5\right]$ are shown.
Not surprisingly, a wealth of modes is present in the parameter space.
In general, the resonance frequencies decrease when deformation is increased, as discussed above.

\begin{figure}
    \centering

    \begin{subfigure}{1.\columnwidth}
        \includegraphics[width=1.\textwidth]{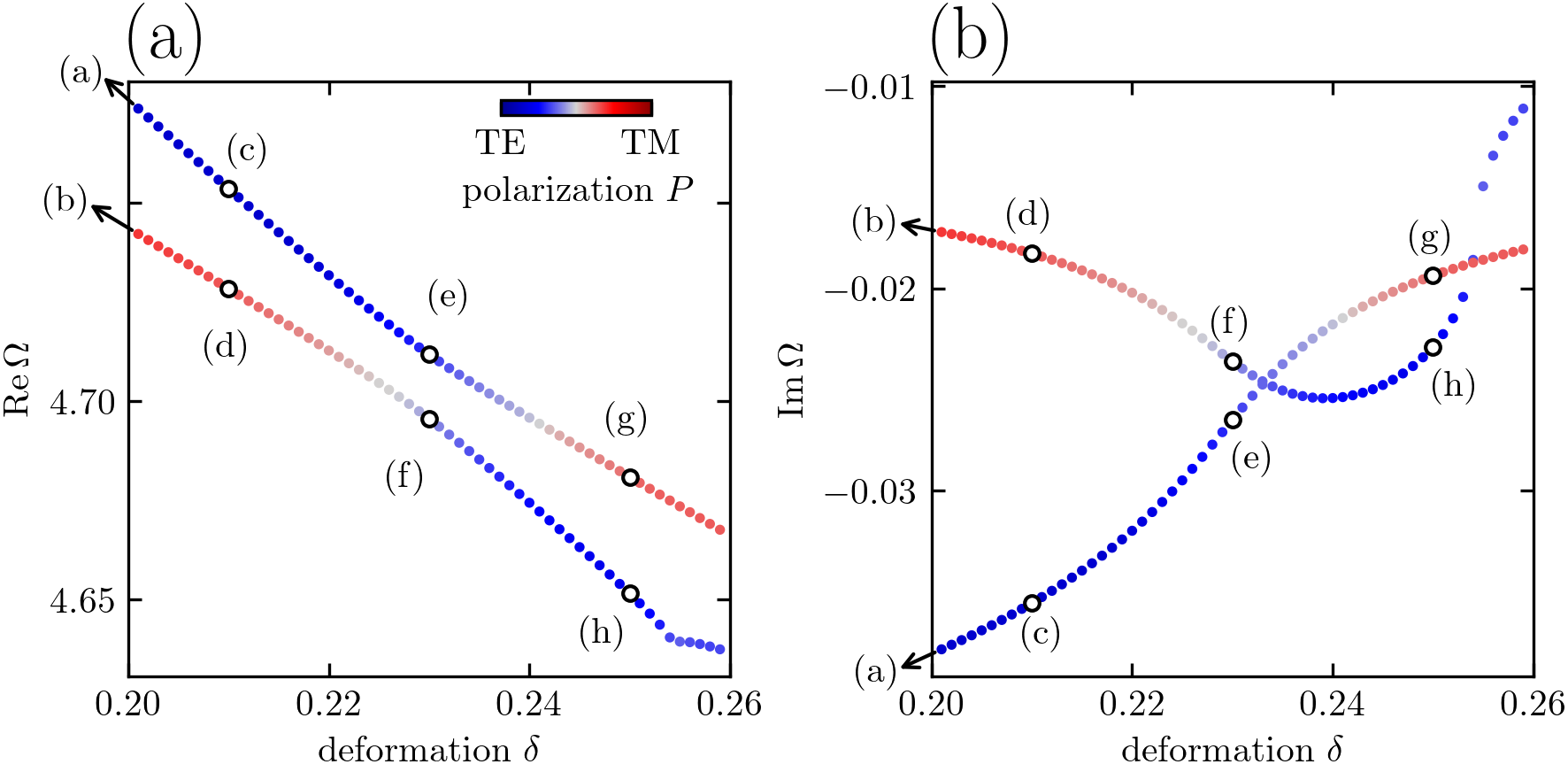}
    \end{subfigure}
    
    \hfill
    \begin{subfigure}{.45\columnwidth}
        \includegraphics[width=1.\textwidth]{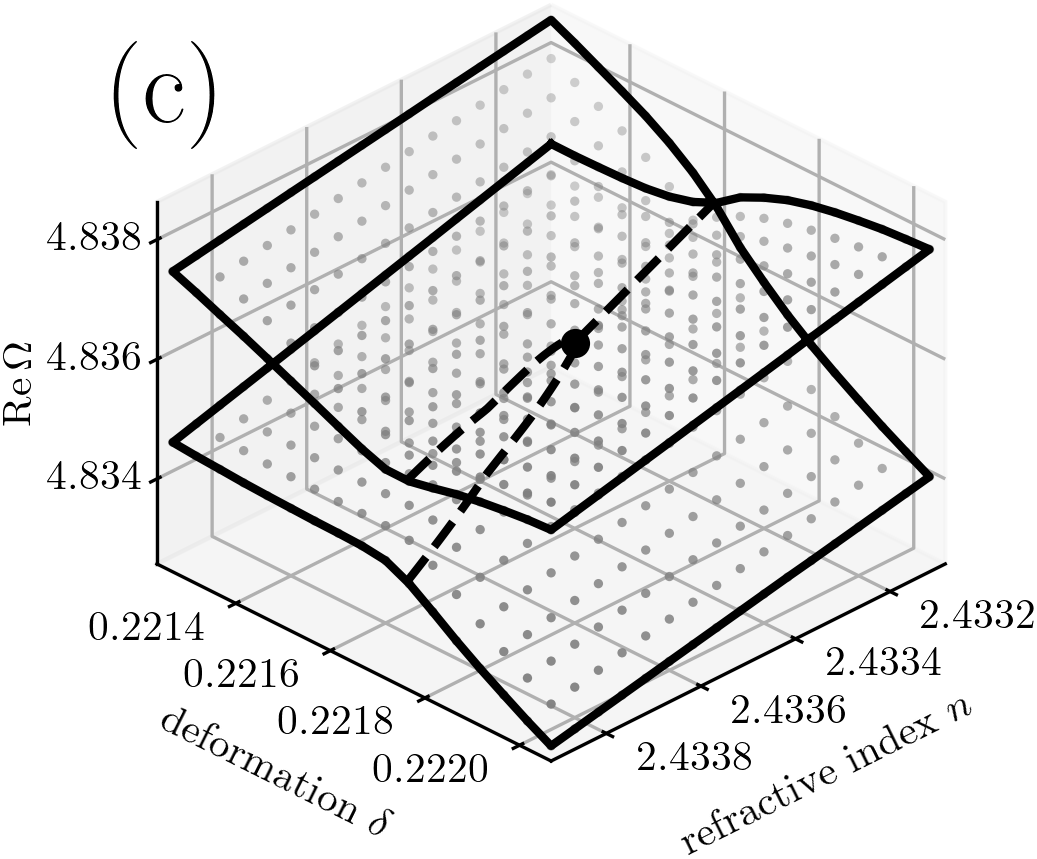}

        \addtocounter{subfigure}{2}
    \end{subfigure}\hspace{1em}
    \begin{subfigure}{.45\columnwidth}
        \includegraphics[width=1.\textwidth]{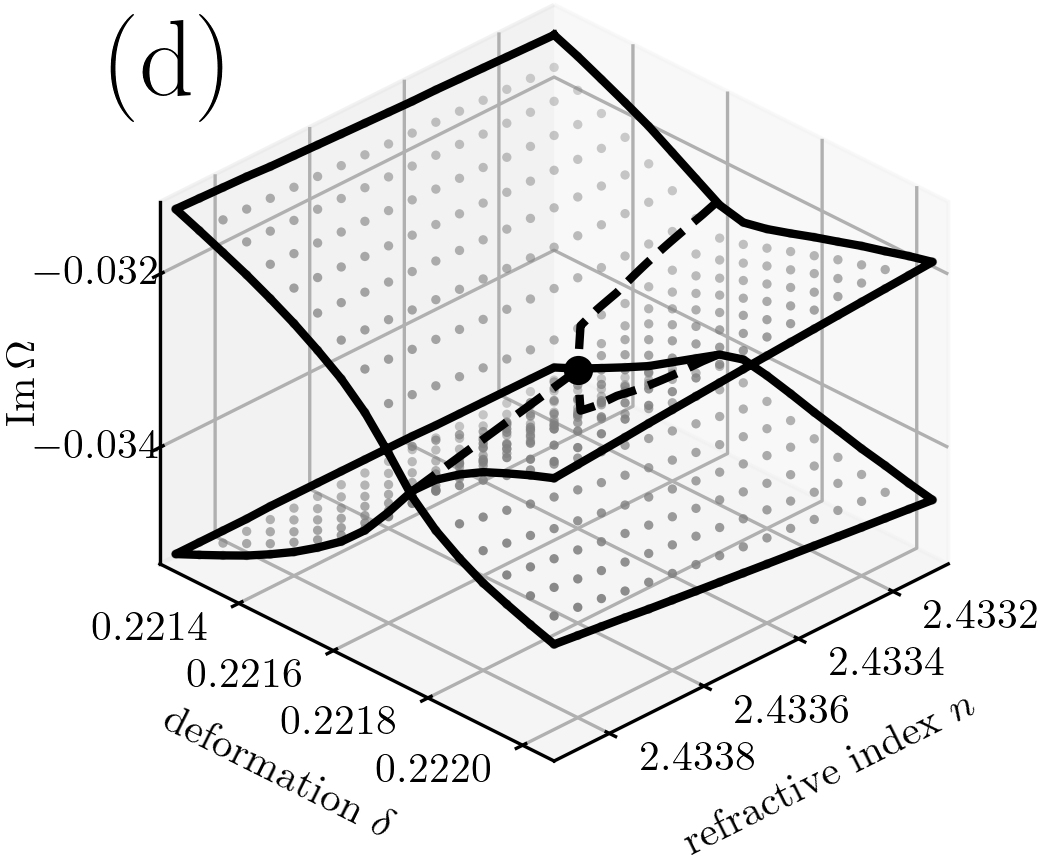}
    \end{subfigure}
    
    \caption{
    Eigenvalue evolution of $(2,6)$-TM and $(2,2)$-TE modes.
    The (a) real and (b) imaginary components of the eigenfrequency, $\Omega$.
    The annotated letters relate to Fig.~\ref{fig:cone:arc_fields} (\ref{fig:cone:husimi}) for the field (Husimi) functions at the respective parameters.
    (c,d) Riemann surface around the EP, which governs the coupling between the two modes for (c) $\re \Omega $ and (d) $\im \Omega$.
    The gray dots correspond to data  used to calculate the energy landscape.
    The black dots denote the positions of the EP, which is located at ${\cyldeform \approx 0.2217}$ and ${n \approx 2.4335}$.
    The black solid lines surrounding the data points illustrate the non-trivial shapes of the surfaces.
    The black dashed lines denote the branch cuts with $\re \Omega \, (\im \Omega) = 0$.
    }

    \label{fig:cone:arc_riemann}
\end{figure}

We are interested in the coupling of TM and TE polarized modes.
Coupling can only occur if the complex eigenvalues, $\Omega$, 
if two modes are similar for certain $\cyldeform$.
This can be understood using an effective Hamiltonian $H$ to describe the system, where
\begin{equation}\label{eq:cone:schroedinger}
    \hamiltonian \psi = \Omega \psi \: . 
\end{equation}
In the case of two coupled modes, the Hamiltonian is a $2\times2$ matrix
\begin{equation}\label{eq:cone:eff_hamiltonian}
    \hamiltonian = \begin{pmatrix}
        \Omega^{(1)} & V \\
        W & \Omega^{(2)}
    \end{pmatrix}, \quad 
\end{equation}
where $\Omega^{(1,2)}$ are the frequencies of the uncoupled modes and $V,W$ are the coupling coefficients.
The two modes can only interact meaningfully if $\sqrt{|VW|} \approx |\Omega^{(1)} - \Omega^{(2)}|$.
If this condition is satisfied, an avoided resonance crossing (ARC) occurs in either the real or imaginary component of $\Omega$, implying a level crossing in the other component.
An exceptional point (EP) is present when $V=0$ or $W=0$, which is accompanied by a coalescence of the two resulting eigenvalues and vectors \cite{heiss2000repulsion, heiss2004exceptional}.

Numerous ARCs are visible in Fig.~\ref{fig:cone:spectrum}.
ARCs between TM and TE modes are the most abundant.
They are easily identified by level repulsion and polarization switching of two (or more) resonance branches.
Examples can be found at $(\delta, \re \Omega) \approx (0.25, 4.7)$ or $(0.27, 4.75)$.
TM-TM and TE-TE ARCs also exist, but they are much less prevalent.
A TM-TM avoided crossing occurs at $(\delta, \re \Omega) \approx(0.05, 5.1)$.
TE-TE ARCs are even less common. 
One emerges at $(\delta, \re \Omega) \approx (0.15, 5.15)$.
A third mode with TM polarization is involved in the ARC, which implies an underlying exceptional point of order 3, as three modes participate in mode coupling.
At an EP of third order, three modes (and their eigenvalues) are collinear. 
Higher-order EPs are more sensitive to small perturbations, as the energy level splitting near an $n$-th order EP scales with the $n$-th root \cite{hodaei2017enhanced, wiersig2022response, kullig2023higher}.

One question to explore is why TM-TE ARCs are the most common type. 
First, let us consider the factors that exclude ARCs from occurring between two modes.
Initially, we focus on TM-TM and TE-TE modes with identical radial quantum number, $\qnrr$.
Modes with lower $\qnzz$ tend to concentrate more of the mode in the region of larger radius (see Fig.~\ref{fig:cone:mode_morphology}).  
This leads to monotonically increasing $\re \Omega(\qnzz)$ for a fixed $\cyldeform$ in Fig.~\ref{fig:cone:spectrum}.
This is the case for any $\cyldeform$, hence no crossings can occur.
As a consequence, only modes with different $\qnrr$ can exhibit crossings in the first place.
Thus, different $\qnzz$ are required in order for two resonances to have similar values of $\re \Omega$.
This affects the mode lifetimes $\im \Omega$, which
must also be similar in order to minimize  $|\Omega^{(1)} - \Omega^{(2)}|$.
Due to these two aspects, TM-TM and TE-TE ARCs are very rare in comparison to TM-TE ARCs.
The boundary conditions in Eqs.~(\ref{eq:3d:boundary_conditions}) lead to $\cyldeform$ affecting the frequencies of TM/TE modes differently, thereby creating the possibility for mode interaction.

\begin{figure}[!htbp]
    \centering

    \begin{tblr}{c|[dashed]c|[dashed]c}
        &
        upper branch &
        lower branch \\
        \hline[dashed]
        \rotatebox[origin=c]{90}{$\cyldeform = 0$} & 
        \begin{minipage}{.39\columnwidth}
            \includegraphics[width=1.\textwidth]{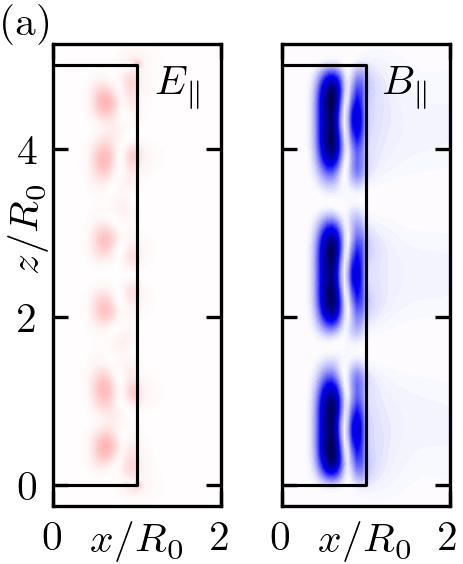}
        \end{minipage} & 
        \begin{minipage}{.39\columnwidth}
            \includegraphics[width=1.\textwidth]{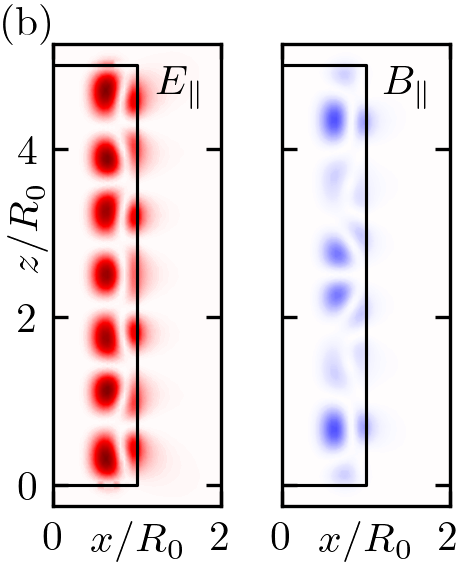}
        \end{minipage} \\
        \hline[dashed]
        \rotatebox[origin=c]{90}{$\cyldeform = 0.21$} & 
        \begin{minipage}{.39\columnwidth}
            \includegraphics[width=1.\textwidth]{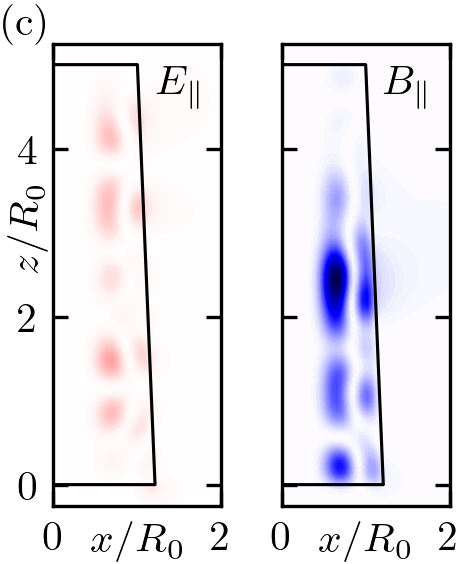}
        \end{minipage} & 
        \begin{minipage}{.39\columnwidth}
            \includegraphics[width=1.\textwidth]{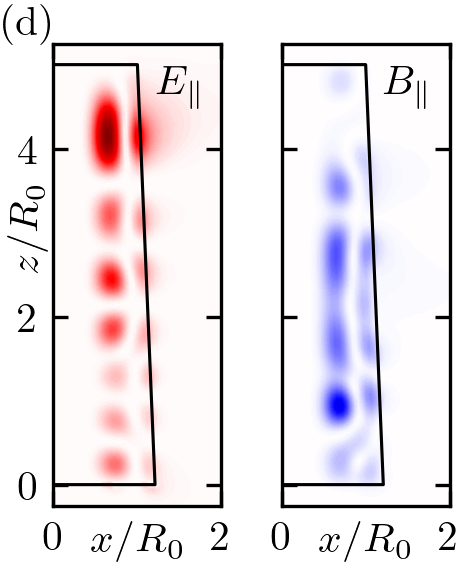}
        \end{minipage} \\
        \hline[dashed]
        \rotatebox[origin=c]{90}{$\cyldeform = 0.23$} & 
        \begin{minipage}{.39\columnwidth}
            \includegraphics[width=1.\textwidth]{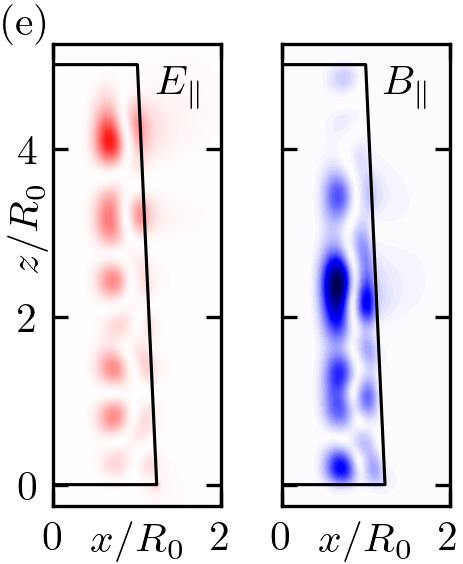}
        \end{minipage} & 
        \begin{minipage}{.39\columnwidth}
            \includegraphics[width=1.\textwidth]{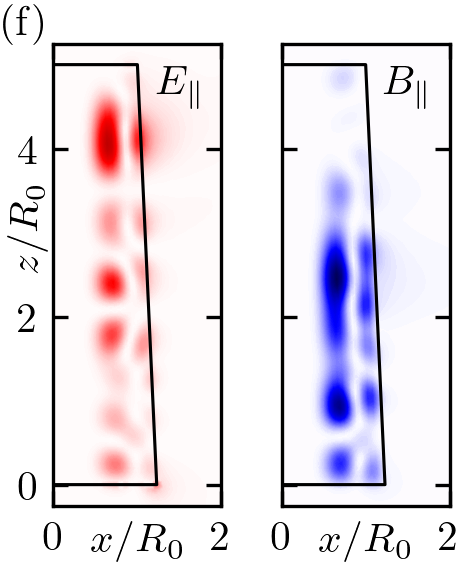}
        \end{minipage} \\
        \hline[dashed]
        \rotatebox[origin=c]{90}{$\cyldeform = 0.25$} & 
        \begin{minipage}{.39\columnwidth}
            \includegraphics[width=1.\textwidth]{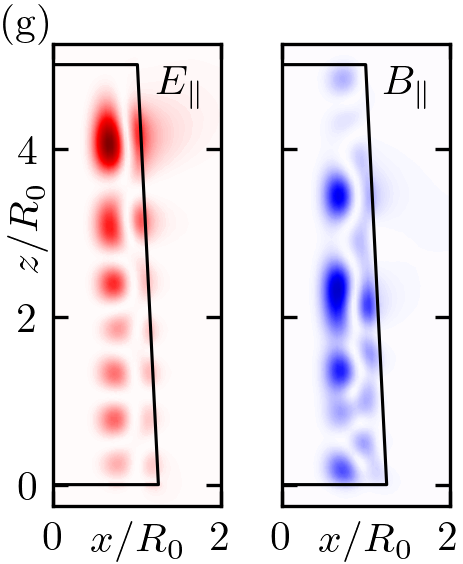}
        \end{minipage} & 
        \begin{minipage}{.39\columnwidth}
            \includegraphics[width=1.\textwidth]{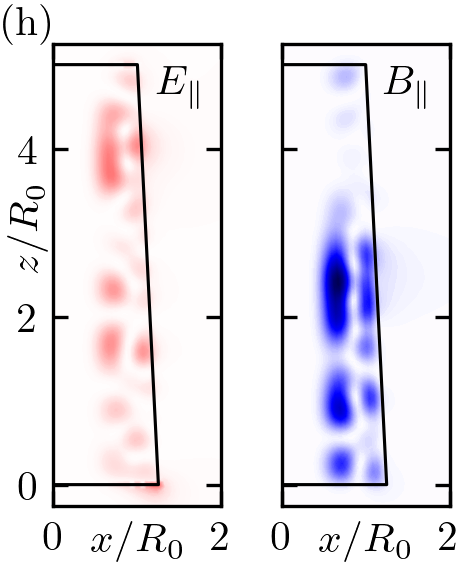}
        \end{minipage} 
    \end{tblr}
    
    \caption{
    Field functions of (2,6)-TM and (2,2)-TE modes for (a,b) $\cyldeform = 0$ and (c-h) at the ARC (see Fig.~\ref{fig:cone:arc_riemann}(a,b)).
    The fields are normalized so that $\max |E_j|^2 = 1$ and $\max |B_j|^2 = 1$, where the subscript $j$ denotes the three field components.
    }
    \label{fig:cone:arc_fields}
\end{figure}

In the following, we concentrate on a specific ARC between a $(1,6)$-TM mode and a $(1,2)$-TE mode, as it is the most common ARC type.
Another ARC, where one TM and two TE modes are involved, is shown in the supplementary material in Appendix~\ref{sec:appendix:additional_arc}.
The eigenvalue evolution of the two modes close to the ARC is shown in Fig.~\ref{fig:cone:arc_riemann}(a,b).
First consider $\re \Omega$ in Fig.~\ref{fig:cone:arc_riemann}(a).
The two frequency branches approach each other with increasing $\cyldeform$.
They are closest at $\cyldeform \approx 0.23$, where both modes exhibit TE polarization.
For $\cyldeform < 0.23$ the upper (lower) branch is TE (TM) polarized.
They switch polarization for $\cyldeform > 0.23$.
The imaginary component in Fig.~\ref{fig:cone:arc_fields}(b) shows a level crossing.
This is expected according to the effective Hamiltonian in Eq.~\eqref{eq:cone:eff_hamiltonian}.
When an (anti-)crossing in (real) imaginary components is present, the coupling is  called strong coupling.
A second crossing also occurs in the imaginary part at $\cyldeform \approx 0.255$ that is not related to the ARC being considered, but another ARC that is adjacent.

ARCs indicate the proximity to an EP, where eigenvalues and eigenfunctions coalesce.
To access the EP a second parameter needs to be varied.
We choose the refractive index, $n$, due to the different influence it has on TM and TE modes because of their varying boundary conditions.
Note that an EP, despite having a measure zero, can nevertheless be identified due to the non-trivial topology of the energy landscape surrounding it. %

The landscape, i.e., the Riemann surface, close to the EP of the $(1,6)$-TM and $(1,2)$-TE modes is shown in Fig.~\ref{fig:cone:arc_riemann}(c,d).
The solid black outlines show that the EP needs to be encircled twice in order to return to the starting position. %
The dashed lines show where $\re (\im) \Delta \Omega = 0$, indicating the switch from strong ($\Delta \im \Omega = 0$) to weak coupling ($\Delta \re \Omega = 0$)
when decreasing the refractive index. \cite{heiss2000repulsion, yi2019non}
In the strong coupling regime the parameters in Eq.~\eqref{eq:cone:eff_hamiltonian} are such that $2 \sqrt{|VW|} > |\im (\Omega^{(1)} - \Omega^{(2)})|$, which results in the level crossing occurring in $\re \Omega$. \cite{heiss2000repulsion, yi2019non}
This is the case for $n > 2.4335$ in (see Figs.~\ref{fig:cone:arc_riemann}(a,c)) for the considered TM-TE mode pair.
The weak coupling regime with $2 \sqrt{|VW|} < |\im (\Omega^{(1)} - \Omega^{(2)})|$ is present for $n < 2.4335$ (see Fig.~\ref{fig:cone:arc_riemann}(d)).

The polarization changes are visible in the wave functions in Fig.~\ref{fig:cone:arc_fields}.
The upper branch for $\cyldeform=0$ in Fig.~\ref{fig:cone:arc_fields}(a)  is distinctly TE polarized, as the out-of-plane component of the magnetic field dominates the mode structure.
The lower branch is TM polarized, see Fig.~\ref{fig:cone:arc_fields}(b).
When $\cyldeform$ is increased, the change in mode morphology takes place as discussed in Sec.~\ref{sec:cone:morphology}.
The intensity distribution of the TE mode shifts to lower $z$ where the radius increases due to its low $\qnzz$ (see Fig.~\ref{fig:cone:arc_fields}(c)).
The TM mode has an intensity maximum close to the top due to its high $\qnzz$ (see Fig.~\ref{fig:cone:arc_fields}(d)).
Near the ARC in Figs.~\ref{fig:cone:arc_fields}(e,f)  both $E_\parallel$ and $B_\parallel$ are of significant intensity.
The TM and TE modes hybridize, which causes both TM and TE components being present.
For larger deformations (see Figs.~\ref{fig:cone:arc_fields}(g,h)) the polarization of the two branches switches.

It is evident that EPs govern the coupling between TM and TE modes.
The mode pairs whose crossings do not show ARC behavior in Fig.~\ref{fig:cone:spectrum} could perhaps be coupled if an appropriate deformation were chosen.
Their coupling is also governed by EPs, even though their signatures are not visible here.
Note that the ARC shown here is just one example.
Numerous other avoided crossings are occurring in the examined parameter space.
This includes triplet-mode interaction, an example of which is shown in Appendix~\ref{sec:appendix:additional_arc}.
The underlying EP of third-order could be accessed by varying a third parameter, such as the cavity height, which has proven a useful parameter to control mode coupling in acoustic cavities \cite{igoshin2024exceptional}.

\subsection{3D Husimi functions for interacting TE-TM modes}

\begin{figure}
    \centering

    \begin{tblr}{c|[dashed]c|[dashed]c}
        &
        upper branch &
        lower branch \\
        \hline[dashed]
        \rotatebox[origin=c]{90}{$\cyldeform = 0$} & 
        \begin{minipage}{.39\columnwidth}
            \includegraphics[width=1.\textwidth]{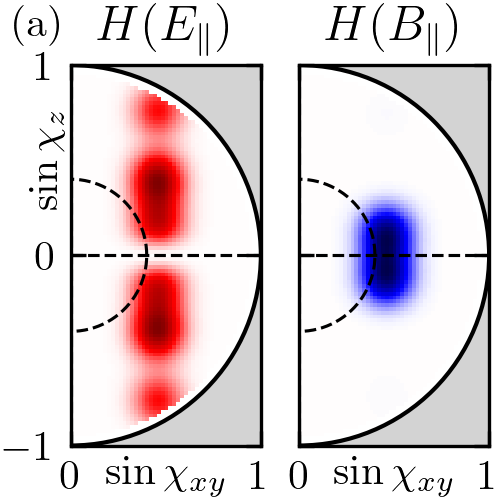}
        \end{minipage} & 
        \begin{minipage}{.39\columnwidth}
            \includegraphics[width=1.\textwidth]{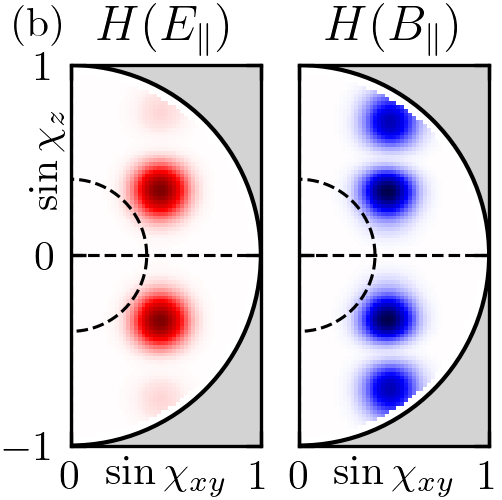}
        \end{minipage} \\
        \hline[dashed]
        \rotatebox[origin=c]{90}{$\cyldeform = 0.21$} & 
        \begin{minipage}{.39\columnwidth}
            \includegraphics[width=1.\textwidth]{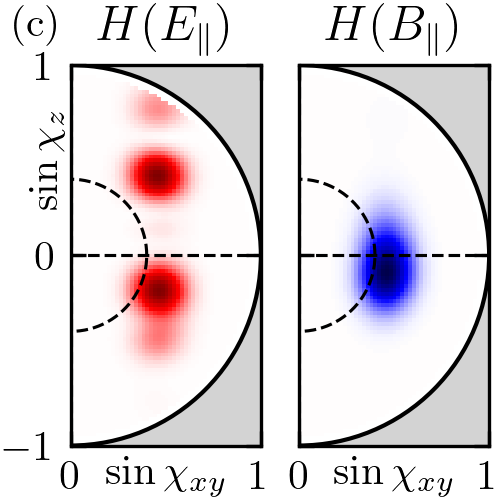}
        \end{minipage} & 
        \begin{minipage}{.39\columnwidth}
            \includegraphics[width=1.\textwidth]{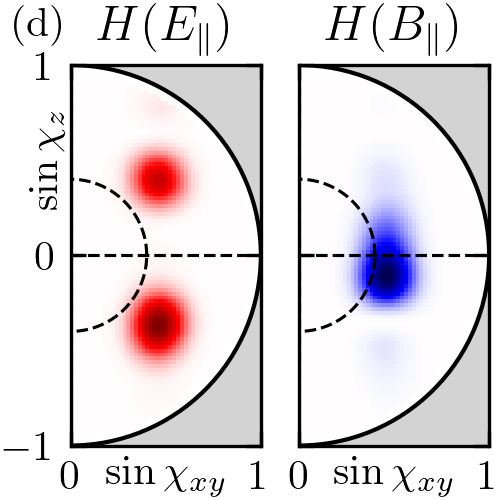}
        \end{minipage} \\
        \hline[dashed]
        \rotatebox[origin=c]{90}{$\cyldeform = 0.23$} & 
        \begin{minipage}{.39\columnwidth}
            \includegraphics[width=1.\textwidth]{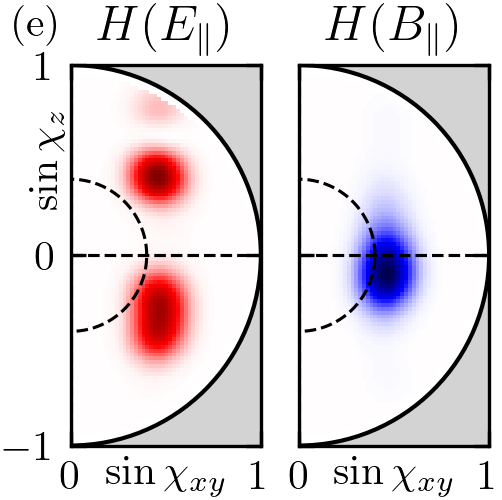}
        \end{minipage} & 
        \begin{minipage}{.39\columnwidth}
            \includegraphics[width=1.\textwidth]{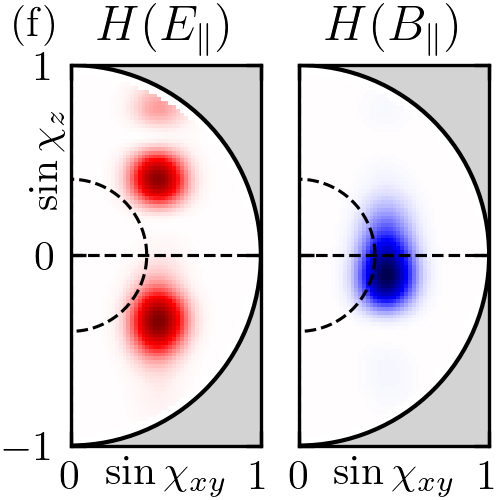}
        \end{minipage} \\
        \hline[dashed]
        \rotatebox[origin=c]{90}{$\cyldeform = 0.25$} & 
        \begin{minipage}{.39\columnwidth}
            \includegraphics[width=1.\textwidth]{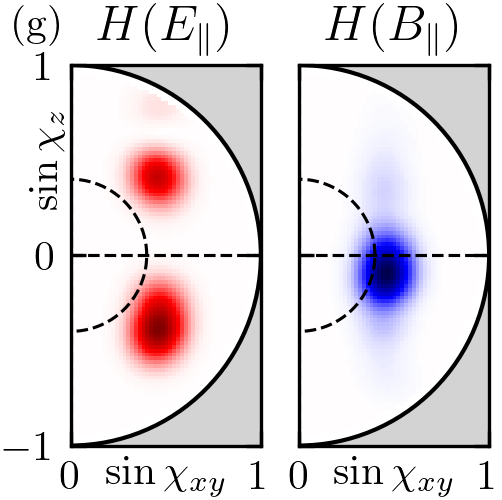}
        \end{minipage} & 
        \begin{minipage}{.39\columnwidth}
            \includegraphics[width=1.\textwidth]{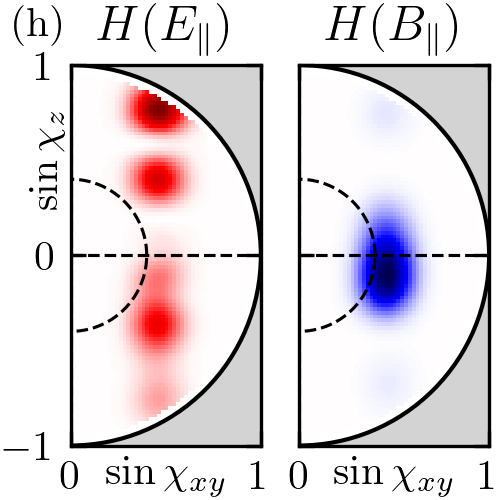}
        \end{minipage} 
    \end{tblr}

    \caption{
    Husimi functions of modes corresponding to ARC (see Figs.~\ref{fig:cone:arc_riemann}(a,b)).
    The red (blue) figures correspond to the $E_\parallel$ ($B_\parallel$) field component used for calculation
    The field functions are shown in Fig.~\ref{fig:cone:arc_fields}, where the numerals in curly brackets refer to the respective parameter.
    }
    \label{fig:cone:husimi}
\end{figure}

The Husimi function signatures of the coupled modes are shown in Fig.~\ref{fig:cone:husimi}.
We choose the Husimi function at $\vec{q}/R_0 = (0,2.5)$, as nearly all wave functions have a nonzero intensity in the vicinity.
The red (blue) colors show the Husimi functions calculated using $E_\parallel$ ($B_\parallel$).
The signatures for the dominant field component of the uncoupled modes at $\cyldeform=0$ are as expected from Fig.~\ref{fig:cone:mode_morphology}.

The dominant out-of-plane component of the TE mode in Fig.~\ref{fig:cone:arc_fields}(a) is the $B_\parallel$ field.
Its Husimi function signature (see Fig.~\ref{fig:cone:husimi}(a)) consists of two intensity maxima, which merge to form one broad intensity maximum. 
This corresponds to the behavior observed in 
Fig.~\ref{fig:cone:mode_morphology}(c) due to the low $\qnzz$ value.
The low-intensity $E_\parallel$ field component consists of small fluctuations due to the TE polarization of the mode, which results in a Husimi function structure resembling a resonance with high $\qnzz$ (compare to Fig.~\ref{fig:husimi:vary_nz}(c)).

The uneven spacing of the intensity maxima in $E_\parallel$ leads to broader peaks in the Husimi function.
The phase space of the TM polarized mode in Fig.~\ref{fig:cone:arc_fields}(b) (left panel) shows the typical behavior of $\qnzz=6$ in $E_\parallel$, as it corresponds to the behavior in Fig.~\ref{fig:husimi:vary_nz}(c). 
Due to the small intensity fluctuations in the $B_\parallel$ component (right panel) the Husimi function yields four intensity maxima.

In Figs.~\ref{fig:cone:husimi}(c,d), the phase space signatures and the mode morphologies change according to Fig.~\ref{fig:cone:mode_morphology}.
Furthermore, both the $E_\parallel$ and $B_\parallel$ components become significant near the ARC.
This is seen, for example, in Figs.~\ref{fig:cone:arc_fields}(e,f) and \ref{fig:cone:husimi}(e,f): The electric fields and phase space representations of both eigenmodes display TM mode characteristics, while the magnetic field (and its Husimi function) show TE behavior.
The mode hybridization caused by the coupling is thus expressed in real and in phase space.
When the system moves away from the ARC, the polarization of the modes increases. 
The Husimi function signatures in Figs.~\ref{fig:cone:husimi}(g,h) reflect the behavior in Figs.~\ref{fig:cone:husimi}(c,d).
Now the polarization is switched, which is reflected in phase space.
For fields and Husimi functions at the EP we refer to  Appendix~\ref{sec:appendix:husimi_at_ep}.

\section{Conclusion}
\label{sec:conclusion}

We have extended the Husimi function formalism to  
3D cavities.
For cylindrical cavities, we found robust signatures that meaningfully reflect mode properties such as quantum numbers and mode morphology.
These signatures are preserved when perturbations are introduced and intricate resonance structures are present, making the %
Husimi functions a useful tool and feasible to characterize  realistic 3D cavities.

Secondly, we used the this %
formalism to investigate, in detail, the interaction between TM and TE modes crucial in 3D systems. 
In particular, we studied their coupling in slightly conical cavities (slanted cylinders).
We found that the smooth change of the mode character between TE and TM is governed by a network of EPs, structuring the parameter space defined by the deformation $\cyldeform$ and refractive index $n$.
A wealth of mode interactions was found in the examined parameter space.
We found that ARCs are most common for mode pairs involving one TM and one TE mode, as a result of the polarization dependent response of the deformation.

The 3D-Husimi function proved to be a useful tool to analyze the exceptional points and the evolution of  mode morphology across an ARC. 
In the future it can play a significant role in understanding and illustrating the mode morphology in generic 3D cavities, allowing for improved access in experiments.
The Husimi functions reveal the most important phase space components and, equally important, the polarization contributions of the resonances.
Generalized Husimi functions complete the description of 3D optical cavities and can reveal information about the incoming and outgoing electromagnetic waves and reveal information on the far-field emission of 3D cavities. This will be subject of further studies.

Experimentally, our findings could be accessed via a host of platforms, as ARCs between TM and TE-polarized modes are very common in 3D cavities.
For any given refractive index and wavelength, a suitable ARC can be chosen to which the system geometry can be tailored.
The only crucial aspect is, that the contributing resonances should have a high $\qnzz$, hence the system needs to have dimensions $h/R_0 > 1$.
The presence of the ARC (or the EP) could be confirmed by measuring Stokes' parameters, as the resonances are only unpolarized in the vicinity of the ARC.
We note that the formalism can also be applied to cavities with nonlinearities, such as microlasers with gain \cite{harayama2011two, kwon2013phase}.

\section{Acknowledgments}

T.S.R. would like to thank the Scientific Computing and Data Analysis Section at OIST for access and support. S.N.C. acknowledges support from the TU Chemnitz Visiting Scholar Program. We thank Takahisa Harayama, Jung-Wan Ryu, Chang-Hwan Yi, Roland Ketzmerick, and Arnd Bäcker for discussions, M.H. in particular for the fruitful exchange at Waseda University and at the Center for Theoretical Physics of Complex Systems, Daejeon. This work was partly supported by OIST Graduate University. S.L. acknowledges support from the Japan Society for the Promotion of Science (JSPS) KAKENHI through Grant-in-Aid for Scientific Research (C) Grant No. 23K04617.

\appendix

\section{Husimi functions for TE modes}
\label{sec:appendix:husimi_vary_nz}

\begin{figure}
    \centering

    \begin{subfigure}{0.31\columnwidth}
        \includegraphics[width=1.\textwidth]{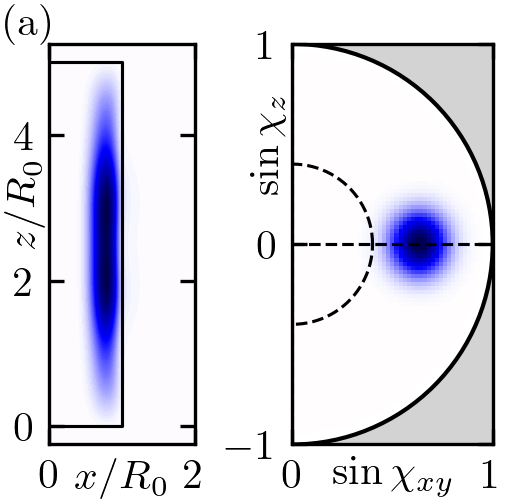}
            
    \end{subfigure}
    \begin{subfigure}{0.31\columnwidth}
        \includegraphics[width=1.\textwidth]{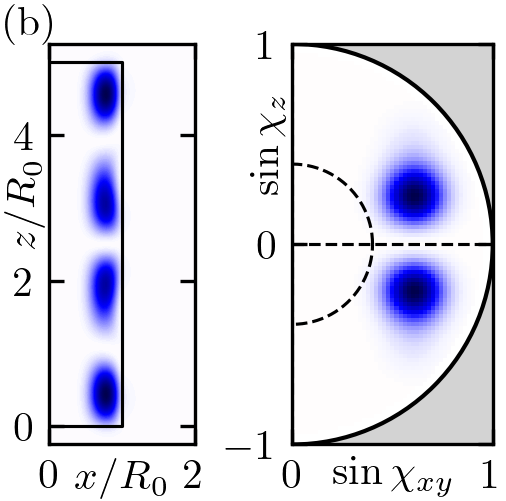}
        
    \end{subfigure}
    \begin{subfigure}{0.31\columnwidth}
        \includegraphics[width=1.\textwidth]{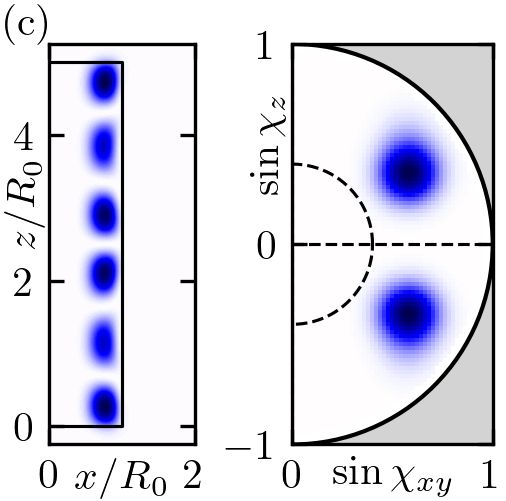}
            
    \end{subfigure}

    \caption{
    Field intensity (left) and Husimi functions (right) for TE modes, with $(\qnrr,\qnphi)=(0,6)$ and various $\qnzz$. 
    The frequencies are (a) $(\qnzz, \Omega) = (0, 3.8377 - \imag \, 0.0070)$, (b) $(\qnzz, \Omega) = (3, 3.9945 - \imag \, 0.0015)$ and (c) $(\qnzz, \Omega) = (5, 4.1901 - \imag \, 0.0006)$
    The phase space functions are shown for $\sin \chi_{xy} > 0$, the rest is given by symmetry.
    }
    \label{fig:app:vary_nz_te}
\end{figure}

The number of intensity maxima in the direction of the symmetry axis ($z$ axis) and its influence on the Husimi functions was discussed for TM modes in Sec.~\ref{sec:husimi:application} (compare Fig.~\ref{fig:husimi:vary_nz}).
For increasing $\qnzz$ the angle of the Husimi function intensity maxima to $\sin \chi_z = 0$ increases monotonically. 
For TE modes the behavior is similar, as demonstrated in Fig.~\ref{fig:app:vary_nz_te}.

For $\qnzz=0$ (see Fig.~\ref{fig:app:vary_nz_te}(a)) only a single intensity maximum is visible in the Husimi function.
For $\qnzz > 0$ the single maximum splits in two maxima, and the angle to $\sin \chi_{xy}$ increases with $\qnzz$, which is shown in Figs.~\ref{fig:app:vary_nz_te}(b-c).

Note that the different boundary conditions for TE polarization results in a mode morphology which is slightly different to that of TM modes (compare Fig.~\ref{fig:husimi:vary_nz}).
This leads to a slight washing out 
of the Husimi function of $\Omega^\mathrm{TE}_3$ mode in Fig.~\ref{fig:app:vary_nz_te}(b), which is not present for its TM mode counterpart in Fig.~\ref{fig:husimi:vary_nz}(b).

\section{Unwrapping the cone}
\label{sec:appendix:cone_mapping}

\begin{figure}[b]

    \centering
    
    \includegraphics[width=.47\columnwidth, valign=m]{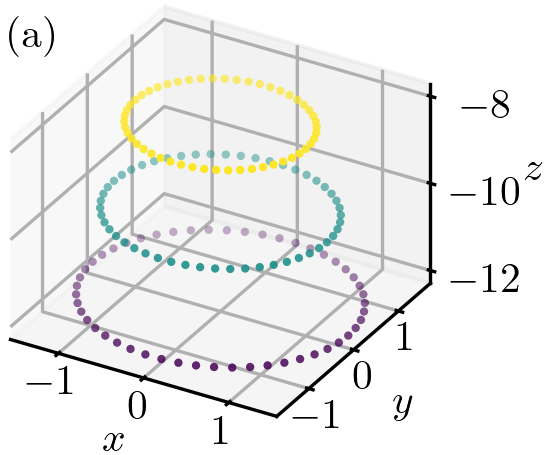}
    \hfill
    \includegraphics[width=.47\columnwidth, valign=m]{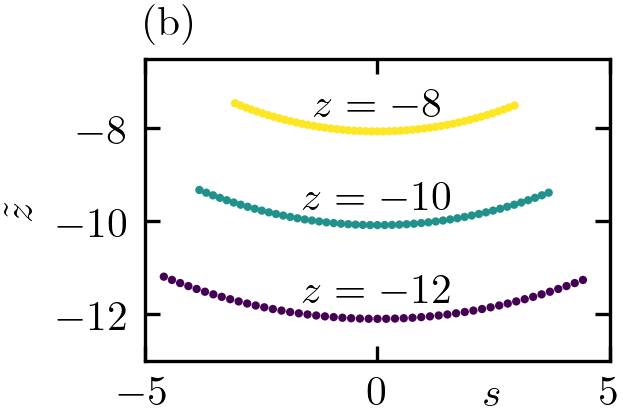}

    \caption{
    (a) Points on the cone surface and (b) the same points mapped to a 2D %
    surface for a cone with opening angle $\alpha = 0.04 \pi$, the spire of which lies in the coordinate origin.
    The color scale corresponds to the $z$ coordinate of the points. 
    }
    \label{fig:app:cone_trafo}
\end{figure}

A cone is described by an opening angle, $\alpha$, and a height, $h_\mathrm{tot}$.
A point on the cone surface can be described by cylindrical coordinates, $(\varphi, z)$, where $\varphi$ is the azimuthal angle and $z$ the cartesian height.
The coordinates can be mapped to a 2D %
surface with the transformation
\begin{gather}
\begin{aligned}
    \varphi' &= \varphi \sin \alpha \quad \text{and} \\
    z' &= z / \cos \alpha, \quad
\end{aligned}
\end{gather}
with an angle scale factor $\sin \alpha$ and a height scale factor $1/\cos \alpha$.
The new coordinates $(\varphi', z')$ can be understood as a 2D %
polar representation of  points on the cylinder surface.
Cartesian coordinates $(s, \tilde{z})$ on the flat cone mantle can then be easily obtained using the usual transformation 
\begin{gather}
\begin{aligned}
    s &= z' \cos \varphi' \quad \text{and} \\
    \tilde{z} &= z' \sin \varphi'. \quad 
\end{aligned}
\end{gather}
An example is shown in Fig.~\ref{fig:app:cone_trafo}, where the transformation is applied to points on the surface of cone with an opening angle of $\alpha = 0.04 \pi$.
The yellow (purple) line corresponds to the boundary of the top (bottom) surface of a slanted cylinder of height $h = 4$ and a deformation $\cyldeform = 0.5$.
Fig.~\ref{fig:app:cone_trafo}(a) shows points on the surface for various $z$ values, mapped to a 2D %
plane in Fig.~\ref{fig:app:cone_trafo}(b).

Using this method, the rotationally symmetric wave functions (such as in Fig.~\ref{fig:cone:mode_morphology}, upper panels) can be mapped to a flat, 2D plane.
A phasor is applied to the numerically calculated 2D %
field function $\psi_m(r,z)$ with angular momentum $m$ such that
\begin{equation}
    \psi(r,z,\varphi) = \psi(r,z) \mathrm{e}^{i m \varphi} \quad .
\end{equation}
The aforementioned transformation is then applied to the field function on the boundary.
The resulting field on the surface for an example of $\Omega_2(\cyldeform=0.1)$ in Fig.~\ref{fig:cone:mode_morphology}(e) is shown in Fig.~\ref{fig:app:husimi_pos_dependence}(a).

\section{Spatial dependence of the Husimi function}
\label{sec:appendix:husimi_spatial_dependence}

\begin{figure}
    \centering

    \includegraphics[width=.8\columnwidth]{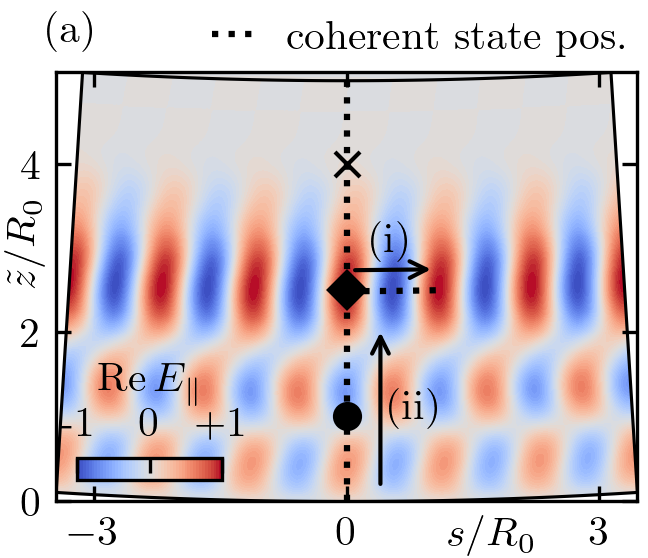}
    
    \includegraphics[width=.8\columnwidth]{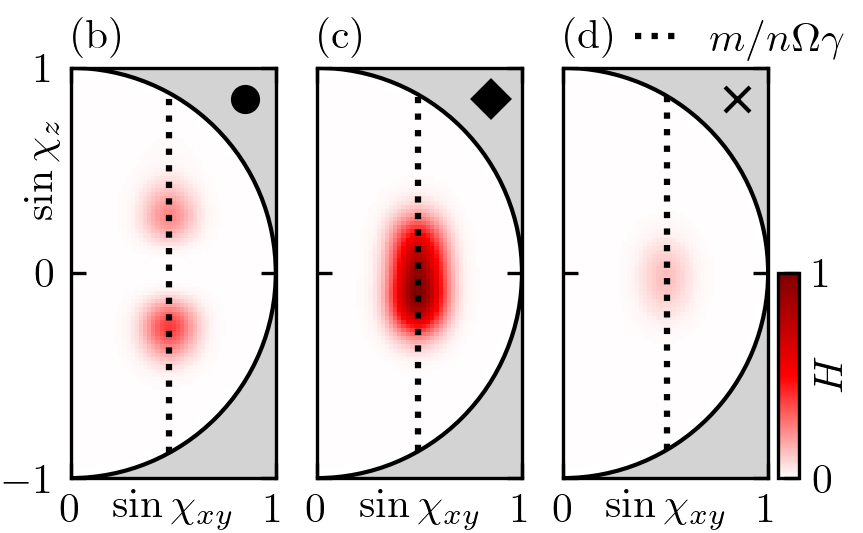}

    \caption{
    (a) Electric field of the $\Omega_2$ mode with $(\qnrr,\qnphi,\qnzz) = (1,6,2)$ of Fig.~\ref{fig:cone:mode_morphology} on the cylinder surface, the black dashed line notes the taken coherent state positions for Fig.~\ref{fig:app:husimi_vary_qz_path2}.
    (b-d) Husimi functions depending on coherent state position.
    The positions are $s=0$ and (b, dot) $z/R_0 = 1$, (c, diamond) $z/R_0 = 2.5$, and (d, cross) $z/R_0 = 4$.
    The black dashed line denotes the expected Husimi function maximum position from Eq.~\eqref{eq:app:husimi_maximumpos}.
    }
    \label{fig:app:husimi_pos_dependence}
\end{figure}

When cavity deformation is increased, the modes lose the mirror symmetry with respect to the $xy$-plane.
The qualitative mode morphology now depends on the location in the cavity which also implies a spatial dependence of the Husimi function.  This is examined in Fig.~\ref{fig:app:husimi_pos_dependence}.
Figure \ref{fig:app:husimi_pos_dependence}a shows the unrolled surface of the $(\qnrr,\qnzz) = (1,2)$ mode.
Three different coherent state positions are selected (see dot, diamond, and cross markers) and the Husimi function is calculated.
The resulting phase spaces are shown in Fig.~\ref{fig:app:husimi_pos_dependence}(b-d).
The Husimi function calculated at the bottom of the cavity (see black dot) shows the signature of a mode with nonzero $\qnzz$, with two distinct Husimi maxima, see Fig.~\ref{fig:app:husimi_pos_dependence}(b).
The phase space in the cavity center (diamond) mainly measures a single broad intensity maximum, which results in a single intensity maximum at $\sin \chi_z \approx 0$, see Fig.~\ref{fig:app:husimi_pos_dependence}(c).
This is a result of the finite size of the coherent state.
The top coherent state (cross) is located in an area where the field intensity is very low.
The field function is dominated by the weak remains of the top intensity maximum, leading to a feint single maximum signature in the Husimi function, see  Fig.~\ref{fig:app:husimi_pos_dependence}(d).

To gain deeper insight into the spatial dependence of the Husimi function we now vary the coherent state position continuously.
We select two different paths on the cavity surface: (i, horizontal dashed line in Fig.~\ref{fig:app:husimi_pos_dependence}(a)) $\tilde{z}=2.5$ with $\varphi\in [0,2\pi/m]$, covering a period in $\euler^{i m \varphi}$ (see Eq.~\eqref{eq:2d:helmholtz_solution}); and (ii, vertical dashed line in Fig.~\ref{fig:app:husimi_pos_dependence}(a)) $\varphi=0$ with $\tilde{z} \in [0,h/\cos \alpha]$.
We only consider the Husimi function values for a fixed value of
\begin{equation}\label{eq:app:husimi_maximumpos}
    \sin \chi_{xy} = \frac{m}{n \re(\Omega) \gamma}
\end{equation}
which is the intensity maximum position.
This expression is familiar from the 2D case \cite{hentschel2002quantum}.
However, it is extended by
\begin{equation}\label{eq:app:husimi_maximumpos_zdependence}
    \gamma = 1 + \cyldeform\left( 1 - \frac{\tilde{z} \cos \alpha}{h}\right)
\end{equation}
which takes the cavity radius and its $z$-dependence $R(z) = R(\tilde{z} \cos \alpha)$ from Eq.~\eqref{eq:cone:radius} into account.

\begin{figure}
    \centering

    \begin{tblr}{c|[dashed]c|[dashed]c}
        & $\cyldeform = 0$ & $\cyldeform=0.1$ \\
        \hline[dashed]
        \rotatebox[origin=c]{90}{$\qnzz = 2$}
        &
        \includegraphics[width=8em, valign=m]{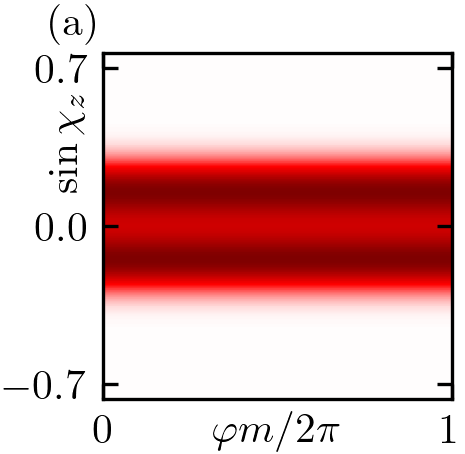}
        &
        \includegraphics[width=8em, valign=m]{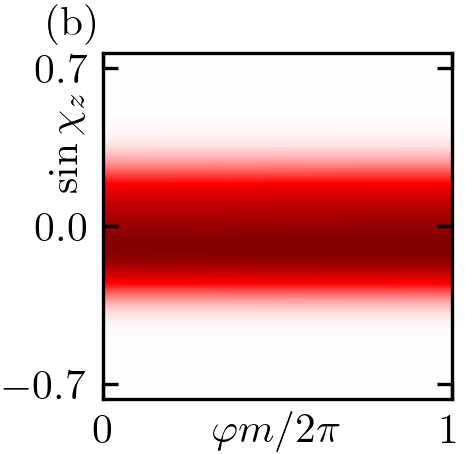}
    \end{tblr}
    
    \caption{
        Husimi functions along path (i) (see Fig.~\ref{fig:app:husimi_pos_dependence}(a)) and $\sin \chi_{xy}$ according to Eq.~\eqref{eq:app:husimi_maximumpos} for modes with $\qnzz = 2$.
        The deformations are (a) $\cyldeform = 0$ and (b) $\cyldeform=0.1$.
        }
    
    \label{fig:app:husimi_vary_qz_path1}
\end{figure}

\begin{figure}
    \begin{tblr}{c|[dashed]c|[dashed]c}
        & $\cyldeform = 0$ & $\cyldeform=0.1$ \\
        \hline[dashed]
        \rotatebox[origin=c]{90}{$\qnzz = 0$} & 
        \begin{minipage}{\spatialwidth}
            \centering
            \includegraphics[width=1.\textwidth]{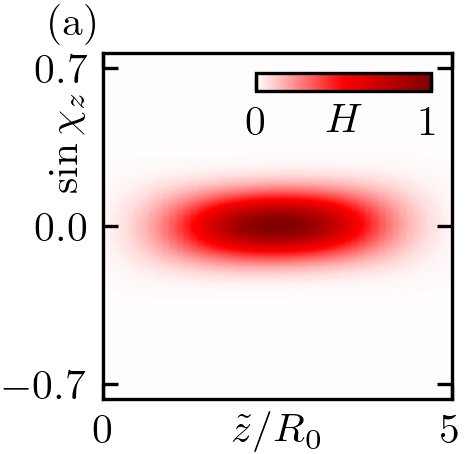}

            $\Omega = 4.820 - \imag \, 0.038$
        \end{minipage} & 
        \begin{minipage}{\spatialwidth}
            \centering
            \includegraphics[width=1.\textwidth]{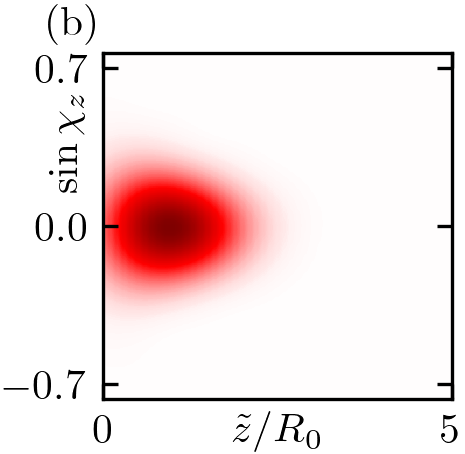}

            $\Omega = 4.487 - \imag \, 0.032$
        \end{minipage}
        \\
        \hline[dashed]
        \rotatebox[origin=c]{90}{$\qnzz = 2$} & 
        \begin{minipage}{\spatialwidth}
            \centering
            \includegraphics[width=1.\textwidth]{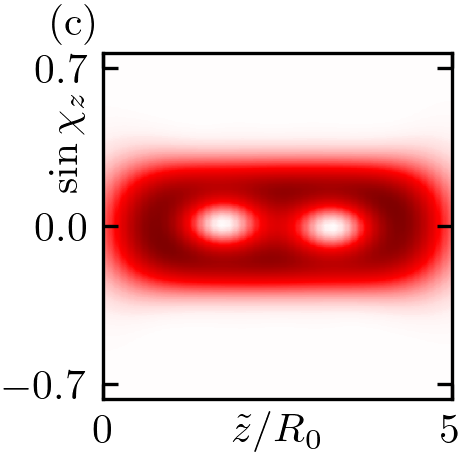}

            $\Omega = 4.854 - \imag \, 0.035$
        \end{minipage} & 
        \begin{minipage}{\spatialwidth}
            \centering
            \includegraphics[width=1.\textwidth]{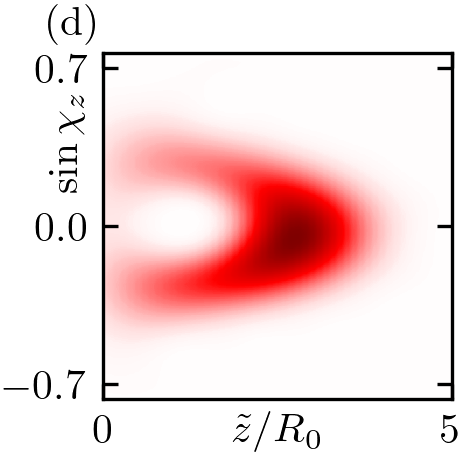}

            $\Omega = 4.630 - \imag \, 0.028$
        \end{minipage}
        \\
        \hline[dashed]
        \rotatebox[origin=c]{90}{$\qnzz = 7$} & 
        \begin{minipage}{\spatialwidth}
            \centering
            \includegraphics[width=1.\textwidth]{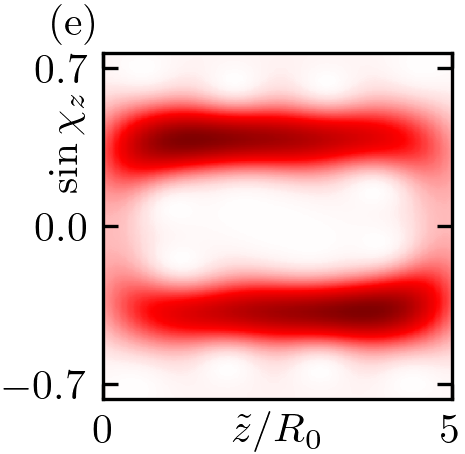}

            $\Omega = 5.185 - \imag \, 0.013$
        \end{minipage} & 
        \begin{minipage}{\spatialwidth}
            \centering
            \includegraphics[width=1.\textwidth]{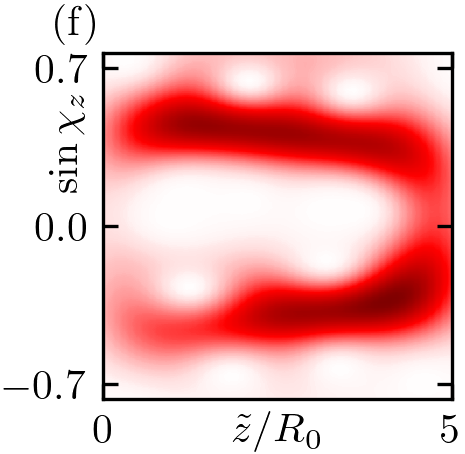}

            $\Omega = 5.002 - \imag \, 0.015$
        \end{minipage}
    \end{tblr}

    \centering
    \caption{
    Husimi functions along path (ii) (cf. Fig.~\ref{fig:app:husimi_pos_dependence}(a)) and $\sin \chi_{xy}$ according to Eq.~\eqref{eq:app:husimi_maximumpos} for modes with $\qnrr=1$ and various $\qnzz$ (cf. Fig.~\ref{fig:cone:mode_morphology}).
    Where the values are taken in real and phase space is shown in Fig.~\ref{fig:app:husimi_pos_dependence}.
    }
    \label{fig:app:husimi_vary_qz_path2}
\end{figure}

The results for path (i) are shown in Fig.~\ref{fig:app:husimi_vary_qz_path1}.
The Husimi function for $\cyldeform=0$ (see Fig.~\ref{fig:cone:mode_morphology}(e)) in Fig.~\ref{fig:app:husimi_vary_qz_path1}(a) does not depend on the azimuthal coordinate on the surface, as is expected for a system with rotational symmetry.
The same is true for the deformed case in Fig.~\ref{fig:app:husimi_vary_qz_path1}(b) (compare Fig.~\ref{fig:cone:mode_morphology}(f)).

Path (ii) yields more interesting results, which are shown in Fig.~\ref{fig:app:husimi_vary_qz_path2}.
The morphology of the mode with $\qnzz=0$ showed a significant change in its morphology when the deformation is increased in Figs.~\ref{fig:cone:mode_morphology}(c,d).
This is also visible in phase space, where the intensitiy maximum shifts from $z\approx h/2$ in Fig.~\ref{fig:app:husimi_vary_qz_path2}(a) to $z\approx h/5$ in Fig.~\ref{fig:app:husimi_vary_qz_path2}(b).
For $\qnzz = 2$ in Figs.~\ref{fig:app:husimi_vary_qz_path2}(c,d) the change in the Husimi function is even more pronounced.
In the unperturbed case, the field intensity maxima positions in real space (compare Fig.~\ref{fig:cone:mode_morphology}(e)) can be identified by the nonzero Husimi function intensities $\sin \chi_z = 0$.
For increased deformation the field intensity maximum at $z \approx h/2$ (cf. Fig.~\ref{fig:cone:mode_morphology}(f)) is clearly visible in the Husimi function as a broad intensity maximum.
At lower height the field is squeezed in $z$ direction, which is visible as a bimodal distribution in the Husimi function, as the real-space checkerboard wave function mimicks a mode with higher $\qnzz$.
For $z>h/2$ the Husimi function signature disappears to to the low field intensity.
When the excitiation in $z$ is high, e.g. $\qnzz=7$, the Husimi function intensity is confined to $\sin \chi_z = \pm 0.4$ in Fig.~\ref{fig:app:husimi_vary_qz_path2}(e).
For the deformed cylinder in Fig.~\ref{fig:app:husimi_vary_qz_path2}(f) the distance in $\sin \chi_z$ of two intensity maxima decreases with $z$.
This is the result of a similar squeeze of the wave function as in Fig.~\ref{fig:app:husimi_vary_qz_path2}(d), but here the maximum at the top is significantly smaller in real space.

Interestingly, the orthogonality constraints on the modes discussed in Sec.~\ref{sec:cone:morphology} are also visible in the Husimi function signatures.
In the deformed cavity, the modes with higher $\qnzz$ occupy different regions in phase space.
The intensity Fig.~\ref{fig:app:husimi_vary_qz_path2}(d) has a distinct minimum at $(z,\sin \chi_z)\approx (2,0)$, which is precisely where the maximum in Fig.~\ref{fig:app:husimi_vary_qz_path2}(b) is located.
The Husimi function for $\qnzz=7$ in Fig.~\ref{fig:app:husimi_vary_qz_path2}(f) appears to encircle the aforementoned phase space structures.

\section{Additional ARCs}
\label{sec:appendix:additional_arc}

\newcommand{\picwidth}{.2\textwidth}

\begin{figure*}
    \centering
    \includegraphics[width=.8\textwidth]{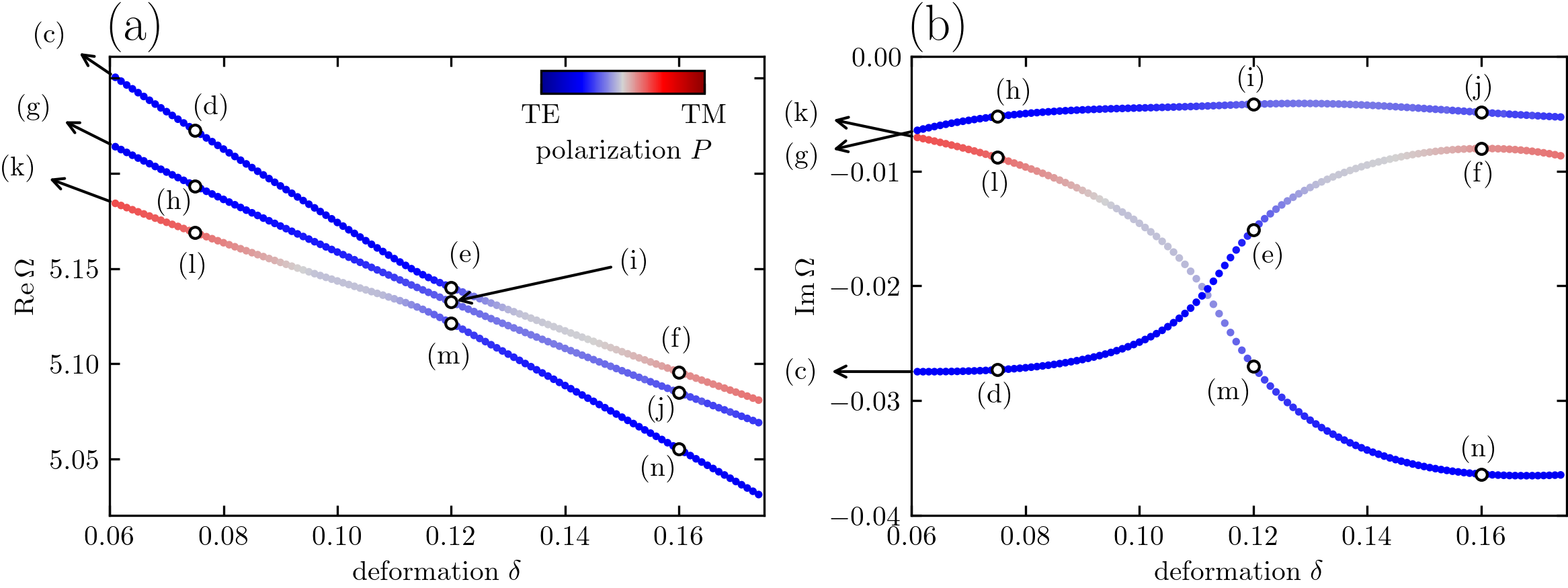}

    \vspace{.5em}

    \begin{tblr}{c|[dashed]c|[dashed]c|[dashed]c|[dashed]c}
        &
        $\cyldeform=0$ &
        $\cyldeform=0.075$  &
        $\cyldeform=0.12$  &
        $\cyldeform=0.16$ 
        \\
        \hline[dashed]
        \rotatebox[origin=c]{90}{upper branch} & 
        \begin{minipage}{\picwidth}
            \includegraphics[width=1.\textwidth]{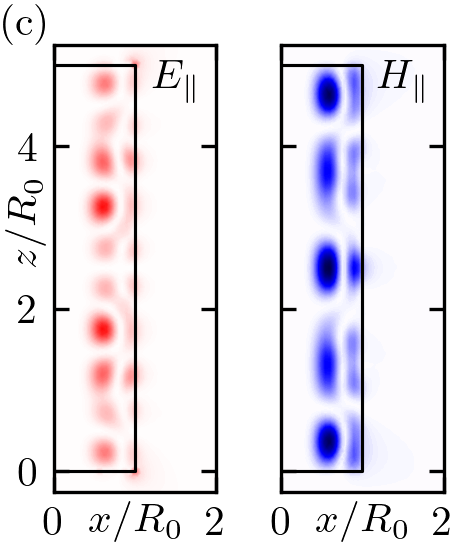}
        \end{minipage} & 
        \begin{minipage}{\picwidth}
            \includegraphics[width=1.\textwidth]{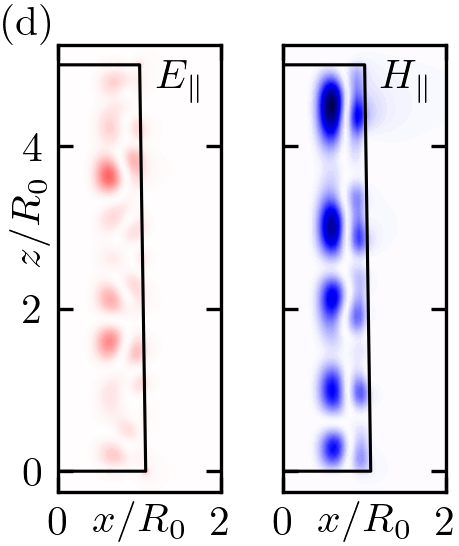}
        \end{minipage} &
        \begin{minipage}{\picwidth}
            \includegraphics[width=1.\textwidth]{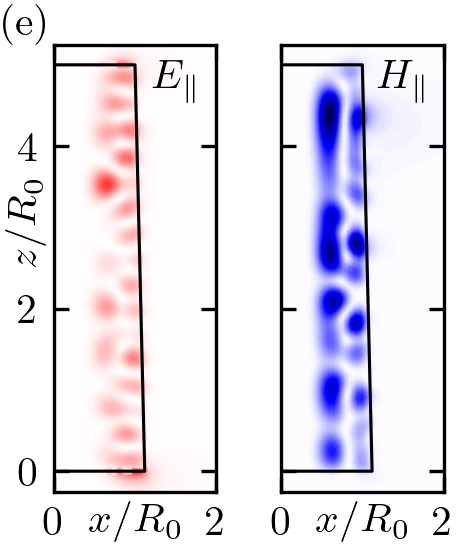}
        \end{minipage} &
        \begin{minipage}{\picwidth}
            \includegraphics[width=1.\textwidth]{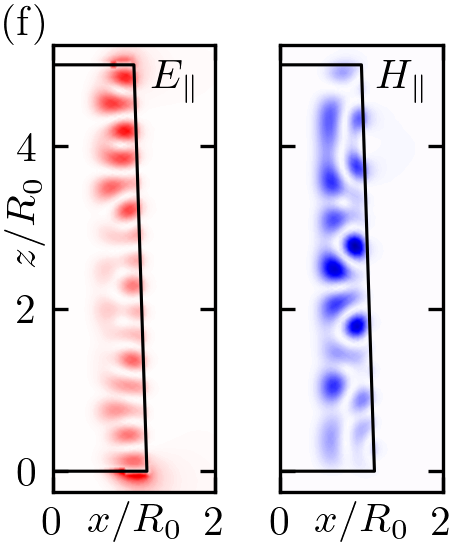}
        \end{minipage}
        \\
        \hline[dashed]
        \rotatebox[origin=c]{90}{center branch} & 
        \begin{minipage}{\picwidth}
            \includegraphics[width=1.\textwidth]{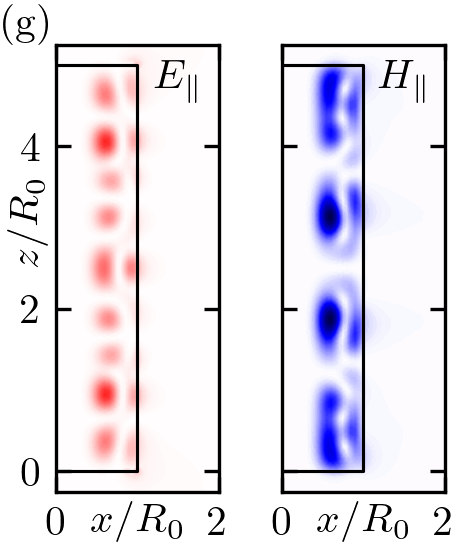}
        \end{minipage} & 
        \begin{minipage}{\picwidth}
            \includegraphics[width=1.\textwidth]{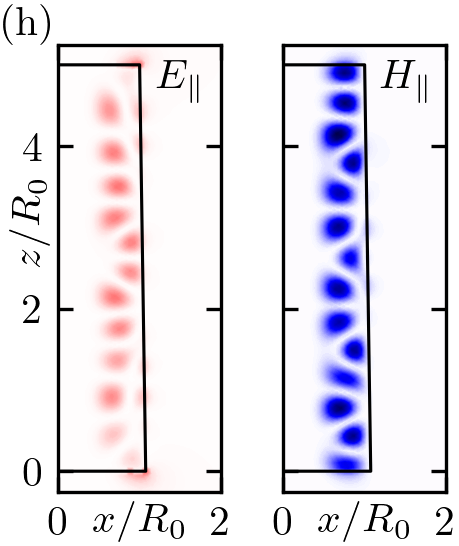}
        \end{minipage} &
        \begin{minipage}{\picwidth}
            \includegraphics[width=1.\textwidth]{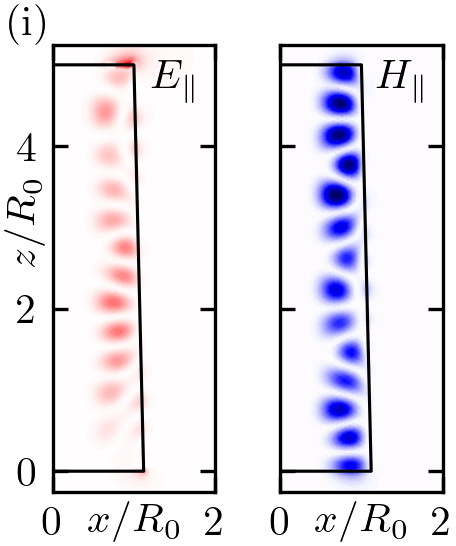}
        \end{minipage} &
        \begin{minipage}{\picwidth}
            \includegraphics[width=1.\textwidth]{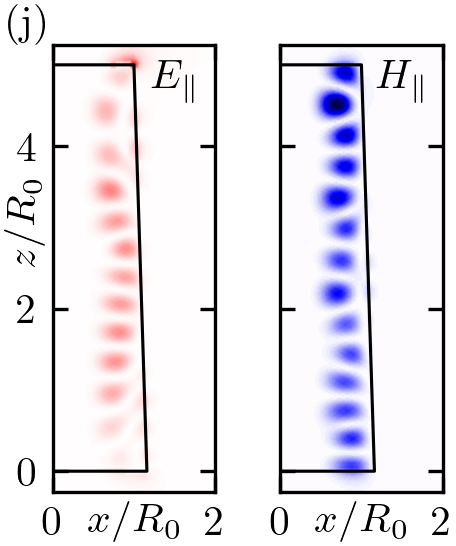}
        \end{minipage}
        \\
        \hline[dashed]
        \rotatebox[origin=c]{90}{lower branch} & 
        \begin{minipage}{\picwidth}
            \includegraphics[width=1.\textwidth]{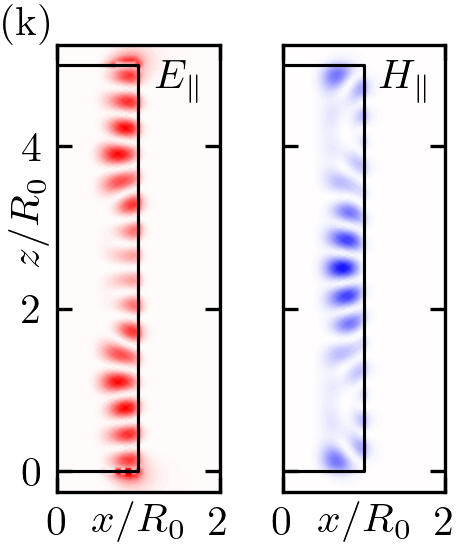}
        \end{minipage} & 
        \begin{minipage}{\picwidth}
            \includegraphics[width=1.\textwidth]{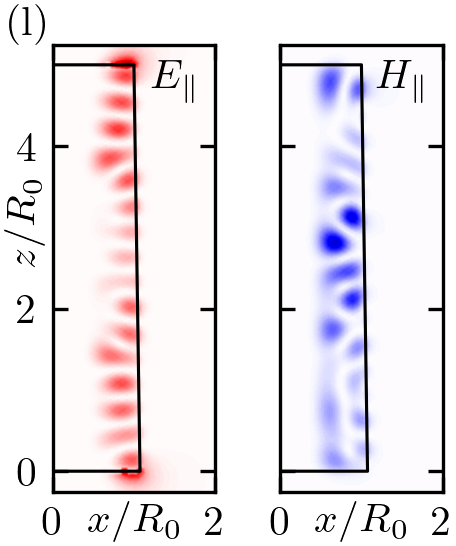}
        \end{minipage} &
        \begin{minipage}{\picwidth}
            \includegraphics[width=1.\textwidth]{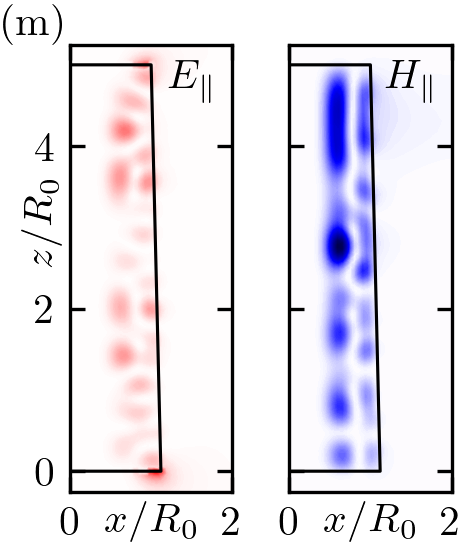}
        \end{minipage} &
        \begin{minipage}{\picwidth}
            \includegraphics[width=1.\textwidth]{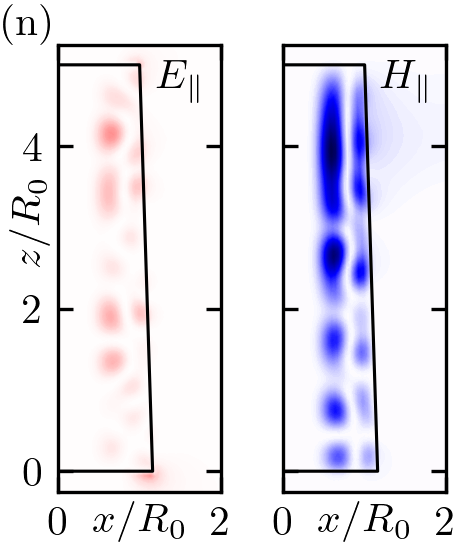}
        \end{minipage}
    \end{tblr}

    \caption{
    (a,b) ARC involving two TE modes and one TM mode and (c-n) wave functions at characteristic points of the ARC.
    The unperturbed frequencies are (c) $\Omega_1 = 5.384- \imag \, 0.026$, (g) $\Omega_2 = 5.326- \imag \, 0.027$ and (k) $\Omega_3 = 5.255 - \imag \, 0.004$.
    }
    \label{fig:app:additional_arc}
\end{figure*}

There are many types of ARCs evident in Fig.~\ref{fig:cone:spectrum}.
The TM-TE type ARC shown in Fig.~\ref{fig:cone:arc_riemann} in the main text is one possible combination of mode polarizations.
Crossings involving more than one mode also exist, one if which is shown in Fig.~\ref{fig:app:additional_arc}.
It involves two TE and one TM polarized modes. 
Noticeably, only the upper and the lower branch exhibit mode hybridization at the ARC at $\cyldeform=0.12$; this is also visible in the crossing of imaginary components of $\Omega$ in Fig.~\ref{fig:app:additional_arc}(b).
The middle branch experiences an enhanced lifetime, which is caused by the mode pattern resembling a mode with higher $\qnzz$ than the unperturbed mode.
Modes with high $\qnzz$ generally have longer lifetimes.
The behavior of an ARC involving $n$ modes is described by an $n\times n$ Hamiltonian \cite{ bergholtz2021exceptional, kullig2023higher}.

\section{Husimi functions at the EP}
\label{sec:appendix:husimi_at_ep}

\begin{figure}
    \centering

    \begin{subfigure}{.3\columnwidth}
        \includegraphics[width=1.\textwidth]{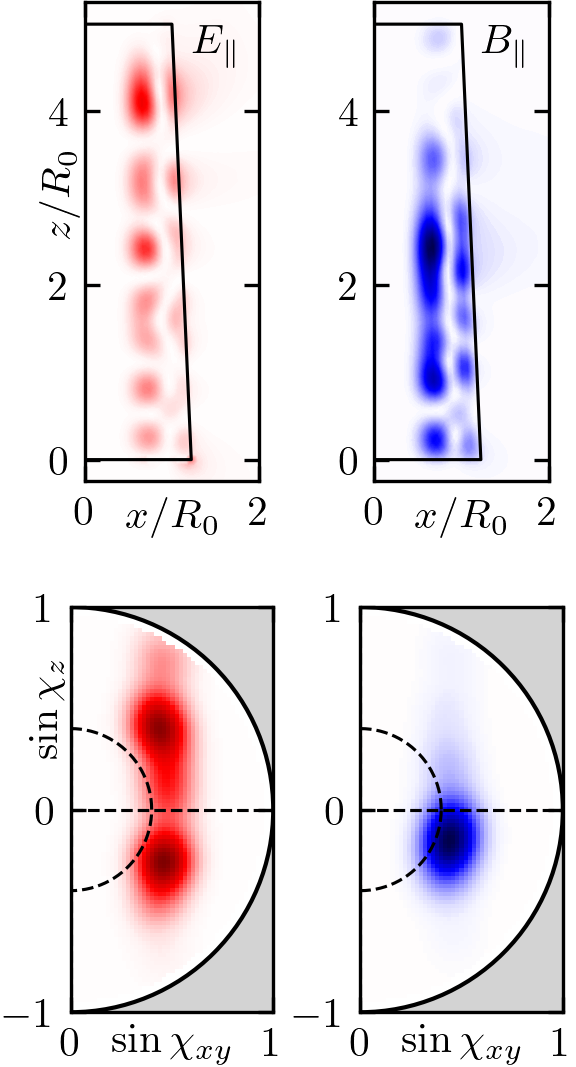}

        \caption{$\Omega_1 = 4.831 - \imag \, 0.033$}
        
    \end{subfigure}\hspace{1em}
    \begin{subfigure}{.3\columnwidth}
        \includegraphics[width=1.\textwidth]{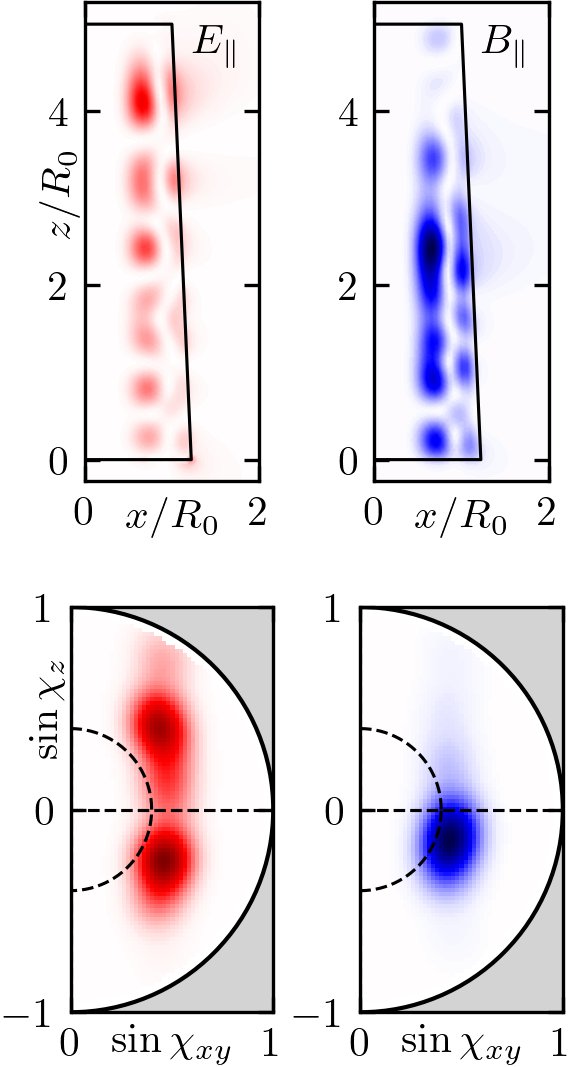}

        \caption{$\Omega_2 = 4.834 - \imag \, 0.033$}
    \end{subfigure}
    
    \caption{
    Top row: Electric (red) and magnetic (blue) fields.  Bottom row: Associated average Husimi functions close to the EP at ${\cyldeform= 0.222}$ and ${n = 2.435}$.
    }
    \label{fig:app:field_and_husimi_close_to_ep}
\end{figure}

The $E$ and $B$-fields, and the Husimi functions close to the EP (see Fig.~\ref{fig:cone:arc_riemann}) are shown in Fig.~\ref{fig:app:field_and_husimi_close_to_ep}.
Coalescence of the mode morphology is clearly visible.
The average Husimi functions $\left\langle H \right\rangle_z$ are shown in the lower panels in Fig.~\ref{fig:app:field_and_husimi_close_to_ep}, where the average is taken over the $z$ position of the coherent state, along a fixed position $s=0$ along at the cavity boundary (compare Fig.~\ref{fig:app:husimi_pos_dependence}).
This shows that the phase space is identical at all points on the cavity surface.

\bibliographystyle{unsrt}
\bibliography{literature}

\end{document}